\documentclass[10pt,english,twocolumn, journal]{IEEEtran}

\usepackage[T1]{fontenc}
\usepackage[latin1]{inputenc}
\usepackage{amsmath,amssymb,amsthm,mathrsfs,amsfonts,dsfont}
\usepackage{geometry}
\usepackage{graphicx}
\usepackage{subcaption}
\usepackage{epstopdf}
\usepackage{cite}
\usepackage{color}
\usepackage{mathtools}
\usepackage{cases}
\usepackage{stackengine}
\usepackage{accents}
\usepackage{lipsum}
\usepackage{soul}

\newcommand\blfootnote[1]{%
  \begingroup
  \renewcommand\thefootnote{}\footnote{#1}%
  \addtocounter{footnote}{-1}%
  \endgroup
}

\newtheorem{thm}{Theorem}[section]
\newtheorem{cor}[thm]{Corollary}
\newtheorem{lem}[thm]{Lemma}
\newtheorem{prop}[thm]{Proposition}

\newtheorem{rem}{Remark}[section]

\renewcommand\appendix{\par
\setcounter{section}{0}
\setcounter{subsection}{0}
\setcounter{figure}{0}
\setcounter{table}{0}
\renewcommand\thesection{{}\Alph{section}}
\renewcommand\thefigure{\Alph{section}\arabic{figure}}
\renewcommand\thetable{\Alph{section}\arabic{table}}
}
\newcommand{\QEDA}{\hfill\ensuremath{\blacksquare}}

\newcommand{\fa}{r_1}
\newcommand{\fb}{r_2}
\newcommand{\fn}{r_n}
\newcommand{\oi}{r_i}

\usepackage{babel}

\allowdisplaybreaks

\begin{document}
\title{Zero-Delay Source-Channel Coding with a Low-Resolution ADC Front End}

\author{Morteza Varasteh$^\dagger$, Borzoo Rassouli$^\dagger$, Osvaldo Simeone$^*$ and Deniz G\"{u}nd\"{u}z$^\dagger$ \\
$^\dagger$ Department of Electrical and Electronic Engineering, Imperial College London, London, U.K.\\
$^*$ Department of Informatics, King's College London, UK.\\
\{m.varasteh12; b.rassouli12; d.gunduz\}@imperial.ac.uk, \thanks{M. Varasteh and B. Rassouli have been supported by the British Council Institutional Links Program under grant number 173605884.

D. Gunduz has received funding from the European Research Council (ERC) through Starting Grant BEACON (agreement No. 677854).} osvaldo.simeone@kcl.ac.uk. \thanks{The work of O. Simeone has received funding from the European Research Council (ERC) under the European Union's Horizon 2020 research and innovation programme (grant agreement No 725731), and was partially supported by the U.S. NSF through grant CCF-1525629.}}

\maketitle
\begin{abstract}
Motivated by the practical constraints arising in emerging sensor network and Internet-of-Things (IoT) applications, the zero-delay transmission of a Gaussian measurement over a real single-input multiple-output (SIMO) additive white Gaussian noise (AWGN) channel is studied with a low-resolution analog-to-digital converter (ADC) front end. Joint optimization of the encoder and the decoder mapping is tackled under both the mean squared error (MSE) distortion and the distortion outage probability (DOP) criteria, with an average power constraint on the channel input. Optimal encoder and decoder mappings are identified for a one-bit ADC front end under both criteria. For the MSE distortion, the optimal encoder mapping is shown to be non-linear in general, while it tends to a linear encoder in the low signal-to-noise ratio (SNR) regime, and to an antipodal digital encoder in the high SNR regime. This is in contrast to the optimality of linear encoding at all SNR values in the presence of a full-precision front end. For the DOP criterion, it is shown that the optimal encoder mapping is piecewise constant and can take only two opposite values when it is non-zero. For both the MSE distortion and the DOP criteria, necessary optimality conditions are then derived for $K$-level ADC front ends as well as front ends with multiple one-bit ADCs. These conditions are used to obtain numerically optimized solutions. Extensive numerical results are also provided in order to gain insights into the structure of the optimal encoding and decoding mappings.\blfootnote{This work was presented in part at the 2016 IEEE International Symposium on Information Theory \cite{ISIT_link_2016}.}

\end{abstract}

\section{Introduction}\label{Ch2_Sec:Introduction}

The power consumed by analog-to-digital converters (ADCs) grows exponentially with the number of bits and linearly with the sampling rate \cite{Walden_1999}, \cite{Murmann_2014}. This technological limitation constrains the resolution of the ADCs used in devices that need to operate under stringent power budgets, such as sensor nodes and mobile devices. As an extreme case, one-bit ADCs are of particular interest, since they can be realized using a simple threshold comparator and without the need for automatic gain control \cite{Park_Simeone_Sahin_Shamai_2014}, \cite{Donnel_Bordersen_2005}.

Motivated by these considerations, the impact of a one-bit ADC front end on the performance of a communication system has been studied in \cite{Singh_Dabeer_Madhow_2009, Koch_Lapidoth_2013, Verdu_2002, Krone_Fettweis_2010, Mezghani_Nossek_2008, Mezghani_Nossek_2007, Jianhua_Heath_2015} for various models, as briefly reviewed below. While these works focus on reliable transmission of digital information over long blocks, in many applications, such as the Internet-of-Things (IoT), cyber-physical systems and wireless sensor networks, the low-delay transfer of analog measurements is crucial. Specific examples include uncoded video transmission techniques \cite{respons_1,respons_2}, memoryless amplify-and-forward relays, wearable sensors that detect neurological impulses \cite{response_5}, and real-time control applications for the IoT  \cite{response_6,response_7}. In light of these emerging applications, this work considers the zero-delay transmission of an analog Gaussian source over a real single-input multiple-output (SIMO) additive white Gaussian noise (AWGN) channel followed by a low-resolution ADC front end. As illustrated in Figure \ref{Ch2_Figure:1}, this problem refers to the transmission of a single sample of a Gaussian source over an individual use of the SIMO AWGN channel. With an infinite resolution front end and equal bandwidth of the Gaussian source and the AWGN channel \cite{Goblick:IT:65}, it is well known that, linear transmission and minimum mean squared error (MMSE) estimation are, respectively, the optimal encoder and decoder under an average power constraint and a mean squared error (MSE) distortion measure. Nonetheless, in the finite resolution scenario studied here, the optimal encoder and decoder mappings have been unknown.

\subsection{Related Works}
Within the spectrum of information theoretic analysis of joint source-channel coding, while the classical Shannon theoretic infinite block-length regime occupies one end of the spectrum, the zero-delay transmission of a single source sample over single channel use lies at the opposite end. Intermediate, but still asymptotic, analyses have also been put forth recently by investigating second-order approximations \cite{response_4,response_9}. For the problem of zero-delay transmission, there is no explicit method to obtain the optimal encoding and decoding mappings, except for the special case of statistically matched source-channel pairs \cite{Gastpar:IT:03}, \cite{Goblick:IT:65}. This special case includes the setting studied in this paper when the receiver front end has infinite resolution. Various solutions have been proposed in the literature for specific unmatched source-channel pairs. Notable examples are the space-filling curves proposed by Shannon \cite{shannon_1949} and Kotelnikov \cite{Kotel'nikov_1959}, and later extended in \cite{Fuldseth_Ramstad_1997,Chung_2000,Vaishampayan_Costa_2003,Ramstad_2002,hekland_floor_ramstad_2009}, for the delay-limited transmission of a Gaussian source over an AWGN channel with bandwidth mismatch between the source and the channel. Other solutions include \cite{Fuldseth_Ramstad_1997, Floor_Ramstad_Wernersson_2007, karlson_skoglund_2010 ,Akyol_Viswanatha_Rose_Ramstad_2014}.

Communication with a finite resolution receiver front end received considerable recent attention, focusing mostly on the channel coding aspects. In \cite{Singh_Dabeer_Madhow_2009}, it is shown that antipodal signalling, or BPSK, is capacity achieving for a real-valued AWGN channel with a one-bit ADC front end, whereas, for the complex counterpart, QPSK is optimal. While these results hold under the assumption that the one-bit ADC is symmetric (that is, it is a zero-threshold comparator), in \cite{Koch_Lapidoth_2013} it is shown that, in the low signal-to-noise ratio (SNR) regime, a symmetric quantizer is not optimal, and the optimal performance is achieved by flash signalling \cite[Def. 2]{Verdu_2002} together with an optimized asymmetric quantizer. In \cite{Krone_Fettweis_2010}, it is shown that, for a point-to-point Rayleigh fading channel with a one-bit ADC front end, under the assumption of perfect channel state information (CSI) at the receiver, QPSK is capacity achieving. Instead, when CSI is not available at the receiver, reference \cite{Mezghani_Nossek_2008} proves that QPSK is optimal above an SNR threshold that depends on the coherence time of the channel, while, for lower SNRs, on-off QPSK achieves the capacity. For the point-to-point multiple-input multiple-output (MIMO) scenarios with a one-bit ADC front end at each receive antenna, the capacity is unknown. In \cite{Mezghani_Nossek_2007}, it is argued that, with perfect CSI at the receiver, QPSK is optimal at very low SNRs, while, in \cite{Jianhua_Heath_2015}, upper and lower bounds on the capacity with perfect transmitter CSI are presented. To the best of our knowledge, joint source-channel coding with a finite resolution ADC is first studied in \cite{ISIT_link_2016}, which is extended to the scenario with a correlated receiver side information in \cite{ITW_varasteh}.

\subsection{Main Contributions and Organization of the paper}
\begin{figure}
\begin{centering}
\includegraphics[scale=0.58]{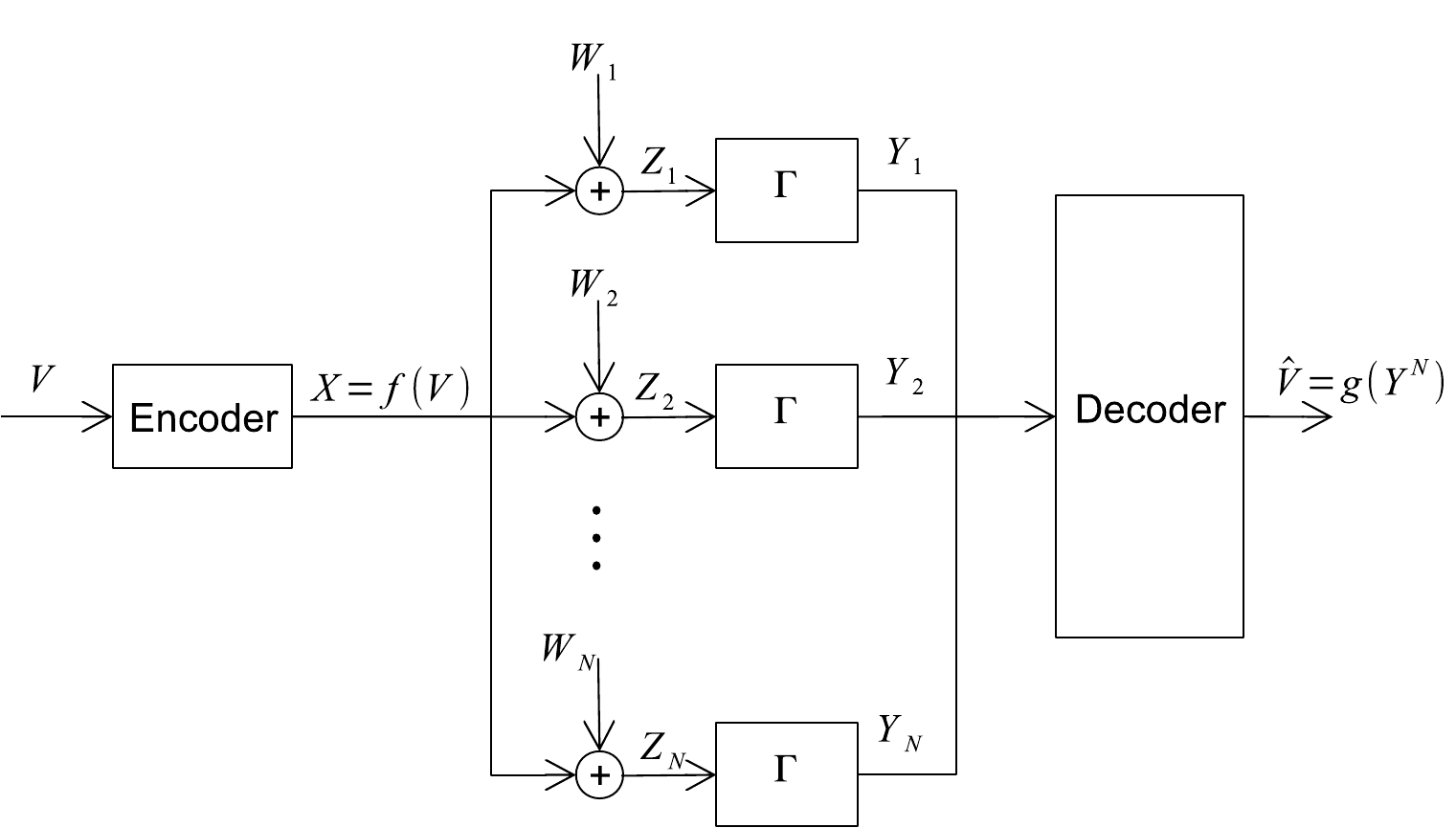}
\caption{System model for the transmission of a single Gaussian source sample over a quantized SIMO AWGN channel with $N$ ADC front ends.}\label{Ch2_Figure:1}
\par\end{centering}
\vspace{0mm}
\end{figure}

We study the optimization of encoding and decoding mappings for the transmission of a Gaussian source sample over the single use of a real SIMO AWGN channel with a low-resolution ADC front end (see Figure \ref{Ch2_Figure:1}). We consider two different criteria, namely the MSE distortion and the distortion outage probability (DOP), with an average power constraint on the channel input. Optimal solutions are derived for the case of a one-bit ADC front end for both criteria. For the MSE distortion, we show that the optimal encoder mapping tends to a linear encoder, which is optimal with a full-precision front end, only in the low-SNR regime. For the DOP criterion, we derive the optimal encoder mapping, showing that it is piecewise constant and that it can take only two opposite values when it is non-zero. For both the MSE distortion and the DOP criteria, we study necessary optimality conditions for $K$-level ADC front ends  as well as for front ends with multiple one-bit ADCs. Extensive numerical results are also provided in order to gain insights into the structure of the optimal encoding and decoding mappings.

The rest of the paper is organized as follows. In Section \ref{Ch2_Sec:System Model}, we introduce the system model. In Section \ref{Ch2_Sec:Average Distortion}, we consider the design of the optimal transceiver under the MSE distortion criterion when the receiver has a single observation of the source. In Section \ref{Ch2_outage-single}, we study the same design problem under the DOP criterion. In Section \ref{Ch2_Sec:Multiple Observation}, we study the more general case in which the receiver makes multiple one-bit observations under both the MSE distortion and the DOP criteria. In Section \ref{Ch2_Sec:Numerical results}, numerical results are provided, followed by the conclusions in Section \ref{Ch2_Sec:Conclusion}.

\textit{Notation}: Throughout the paper, $\mathbb{R}$ denotes the set of real numbers;  uppercase and lowercase letters denote random variables and realizations, respectively. We use $b_{j}^N$ to denote the bit-wise representation of the number $2^N-j,~j=1,\ldots,2^N$ with length $N$. $\mathbb{E}[\cdot]$ and $\mathrm{Pr}(\cdot)$ denote the expectation and probability operators, respectively. Let $f^{'}(x)=\frac{df(x)}{dx}$, $f^{''}(x)=\frac{d^2f(x)}{dx^2}$ denote the first and second order derivatives of the continuously differentiable function $f$ with respect to its argument. The standard normal distribution is denoted by $\mathcal{N}(0,1)$, with cumulative distribution function $\Phi(\cdot)$ and the  complementary cumulative distribution function (CCDF) by $Q(\cdot)$, which is given by
\[Q(z)=\frac{1}{\sqrt{2\pi}}\int\limits_{z}^{\infty}e^{-\frac{t^2}{2}}dt.\] Unless stated otherwise, boundaries of integrals are from $-\infty$ to $\infty$.

\section{System Model}\label{Ch2_Sec:System Model}
We consider the system model in Figure \ref{Ch2_Figure:1}, in which a single sample of a Gaussian source $V\sim\mathcal{N}(0,\sigma_v^2)$ is transmitted over a single use of a quantized SIMO AWGN channel. The encoded signal is given as $X=f(V)$, where $f:\mathbb{R} \rightarrow \mathbb{R} $ is a mapping from the source sample to the channel input, with average transmission power $P=\mathbb{E}[f(V)^2]$. The receiver makes $N$ noisy measurements of the encoded signal, which are digitized by means of a low-resolution ADC front end. Mathematically, each noisy received signal is modelled as
\begin{equation}
Z_i=f(V)+W_i,~i=1,...,N,
\end{equation}
where the noise $W_i\sim \mathcal{N}(0,\sigma_{w_i}^2),~i=1,...,N$, is independent over index $i$. Each received signal $Z_i$ is quantized with a scalar $K$-level ADC producing a quantized signal
\begin{equation}
Y_i= \Gamma (Z_i),~i=1,\ldots,N.
\end{equation}
The scalar $K$-level ADC is characterized by fixed quantization intervals and corresponding quantized levels, namely
\begin{equation}\label{Ch2_Eqn:93}
 \Gamma (z)=y_{(j)},\quad \text{for} \quad z\in\left[z_{(j-1)},z_{(j)}\right),~j=1,\ldots,K,
\end{equation}
where $z_{(j-1)}$ and $z_{(j)}$ are the lower and upper bounds of the interval corresponding to the quantized signal $y_{(j)}$, respectively, for $j=1,\ldots,K$, and we have $z_{(0)}=-\infty$ and $z_{(K)}=\infty.$
Note that the ADCs employed to quantize different channel outputs all have the same quantization intervals and reconstruction levels.

For most of the paper, we will consider a symmetric one-bit ADC with threshold $z_{(1)}=0$ and reconstruction levels $y_{(1)}=1$ and $y_{(2)}=0$:
\begin{equation}\label{Ch2_Eqn:92}
 \Gamma (z)=\left\{\begin{array}{ll}
                 0 & z\geq 0, \\
                 1 & z < 0.
               \end{array}\right.
\end{equation}
We define the SNR as
\begin{align}\label{Ch2_Eqn:75}
\gamma=\frac{NP}{\sum\limits_{i=1}^{N} \sigma_{w_i}^2} .
\end{align}
Based on the quantized signals $(Y_1,...,Y_N)\triangleq Y^N$, the decoder produces an estimate $\hat{V}$ of $V$ using a decoding function $g:\{y_{(1)},\ldots,y_{(K)}\}^N\rightarrow \mathbb{R}$, i.e., $\hat{V}=g(Y^N)$.

Two performance criteria are considered, namely the MSE distortion, which is defined as
\begin{equation}\label{Ch2_Eqn:1}
\bar{D}=\mathbb{E}\left[(V-\hat{V})^2\right],
\end{equation}
and the DOP, which is instead defined as
\begin{equation}\label{Ch2_Eqn:2}
\epsilon(D)=\mathrm{Pr}\left((V-\hat{V})^2\geq D\right).
\end{equation}
In both cases, we aim at studying the optimal encoder function $f$, along with the corresponding optimal estimator $g$ at the decoder, such that $\bar{D}$ and $\epsilon(D)$ are minimized, subject to an  average power constraint. More specifically, as it is common in related works (see, e.g., \cite{Akyol_Viswanatha_Rose_Ramstad_2014}), we consider the unconstrained minimization
\begin{equation}\label{Ch2_Eqn:3}
\begin{aligned}
& \underset{f,g}{\text{minimize}}
& & L(f,g,\lambda),
\end{aligned}
\end{equation}
where
\begin{align}\label{Ch2_Eqn:4}
  L(f,g,\lambda) &=\left\{\begin{array}{ll}
                   \bar{D}+\lambda \mathbb{E}[f(V)^2] & \text{for the MSE criterion}, \\
                   \epsilon(D)+\lambda \mathbb{E}[f(V)^2] & \text{for the DOP criterion},
                 \end{array}\right.
\end{align}
with $\lambda\geq 0$ being a Lagrange multiplier that defines the relative weight given to the average transmission power $\mathbb{E}[f(V)^2]$ as compared to the distortion criterion.

\section{Single Observation: MSE Distortion}\label{Ch2_Sec:Average Distortion}
In this section, we study the design of the encoder and the decoder under the MSE criterion by focusing on the case of a single observation $(N=1)$. For the one-bit ADC in (\ref{Ch2_Eqn:92}), we obtain the optimal encoder and decoder in Section \ref{Ch2_Sec:Optimality Conditions}. Furthermore, we consider the conventional linear transmission and digital modulation schemes for reference in Section \ref{Ch2_Sec:Uncoded transmission} and Section \ref{Ch2_Sec:Digital Transmission}, respectively. Finally, we consider the extensions to a $K$-level front end, and obtain a necessary condition on the optimal mapping in Section \ref{Ch2_Sec:K_level_MSE}. For brevity, throughout this section, we drop the subscript $i=1$ identifying the observation index.

\subsection{Optimal Encoder and Decoder for a One-Bit Front End}\label{Ch2_Sec:Optimality Conditions}
To elaborate on the optimal encoder and decoder for the one-bit ADC in (\ref{Ch2_Eqn:92}), without loss of generality, we write the receiver mapping as
\begin{equation}\label{Ch2_Eqn:5}
g(Y)=\hat{V}=\left\{\begin{array}{ll}
\hat{v}_{(1)}&~Y=0, \\
\hat{v}_{(2)}&~Y=1,
\end{array}\right.
\end{equation}
which is defined by the pair of parameters $(\hat{v}_{(1)},\hat{v}_{(2)})$. In (\ref{Ch2_Eqn:5}), since, for any encoder mapping $f$, the MMSE estimator is optimal under the MSE criterion, and hence also for problem (\ref{Ch2_Eqn:3}), we have $\hat{v}_{(1)}=\mathbb{E}[V|Y=0]$ and $\hat{v}_{(2)}=\mathbb{E}[V|Y=1]$. The next proposition provides the optimal encoder mapping. 

\begin{prop}\label{Ch2_Result:1}
The optimal mapping $f$ for problem (\ref{Ch2_Eqn:3}) under the MSE criterion is unique up to a sign, is an odd function of $v$, and is defined by the implicit equation
\begin{eqnarray}\label{Ch2_Eqn:6}
f(v)e^{\frac{f(v)^2}{2\sigma_w^2}}=\frac{v}{\sqrt{2\pi}\sigma_w\lambda}.
\end{eqnarray}
\end{prop}
\textit{Proof}: See Appendix \ref{Ch2_Appendix:1}.


Illustrations of the optimal mappings satisfying (\ref{Ch2_Eqn:6}) will be given in Section \ref{Ch2_Sec:Numerical results}. Here, we observe that, by expanding the Taylor series of the exponential function in (\ref{Ch2_Eqn:6}), it can be easily verified that, in the low SNR regime, that is, as $\sigma_w^2\rightarrow \infty$, the optimal mapping satisfies the condition $f(v)\varpropto v$, that is, it approaches a linear mapping. Furthermore, given that the optimal mapping $f(v)$ is odd, we can write
\begin{subequations}
\begin{align}
\hat{v}_{(1)}&= \mathbb{E}[V|Y=0]\\
&=\frac{\frac{1}{\sigma_v}\int v\Phi\left(\frac{v}{\sigma_v}\right)\mathrm{Pr}(Y=0|V=v)dv}{\mathrm{Pr}(Y=0)}\\\label{Ch2_Eqn:78}
&=\frac{-2}{\sigma_v}\int v\Phi\left(\frac{v}{\sigma_v}\right)Q\left(\frac{f(v)}{\sigma_w}\right)dv\\\label{Ch2_Eqn:7}
&=\frac{2\sigma_v}{\sqrt{2\pi}}-\frac{4}{\sigma_v}\int\limits_{0}^{\infty}v\Phi\left(\frac{v}{\sigma_v}\right)Q\left(\frac{f(v)}{\sigma_w}\right)dv,
\end{align}
\end{subequations}
and hence the average distortion can be simplified as
\begin{subequations}
\begin{align}\label{Ch2_Eqn:100}
\bar{D}&= \sigma_v^2-\mathbb{E}[V\hat{V}]\\\nonumber
&= \sigma_v^2-\frac{1}{2}\left(\hat{v}_{(1)}\mathbb{E}[V|\hat{V}=\hat{v}_{(1)}]\right.\\\label{Ch2_Eqn:101}
&\left.\quad+\hat{v}_{(2)}\mathbb{E}[V|\hat{V}=\hat{v}_{(2)}]\right)\\\label{Ch2_Eqn:8}
&=\sigma_v^2-\hat{v}_{(1)}^2,
\end{align}
\end{subequations}
where (\ref{Ch2_Eqn:100}) is due to the orthogonality property of MMSE estimation; (\ref{Ch2_Eqn:101}) follows from the fact that the optimal encoder is odd; and (\ref{Ch2_Eqn:8}) is due to the chain of equalities $\mathbb{E}[V|\hat{V}=\hat{v}_{(1)}]=\mathbb{E}[V|Y=0]=\hat{v}_{(1)}=-\hat{v}_{(2)}=-\mathbb{E}[V|\hat{V}=\hat{v}_{(2)}]$.

\subsection{Linear Transmission for One-Bit Front End}\label{Ch2_Sec:Uncoded transmission}
Here we consider the performance of linear transmission in the presence of a one-bit ADC front end. The encoder mapping for linear transmission is given by
\begin{equation}\label{Ch2_Eqn:9}
f(v)=\sqrt{\frac{P}{\sigma_v^2}} v.
\end{equation}
As seen in Section \ref{Ch2_Sec:Optimality Conditions}, linear transmission is asymptotically optimal in the low-SNR asymptotic regime. In the following, we elaborate on its performance for any given channel SNR $\gamma$. The MSE distortion $\bar{D}_l$ achieved by linear transmission, can be found by calculating the integral in (\ref{Ch2_Eqn:7}) with $f(v)$ given in (\ref{Ch2_Eqn:9}) and substituting the result in (\ref{Ch2_Eqn:8}). As a result, the MSE distortion for the linear mapping is obtained as
\begin{eqnarray}\label{dd}
\bar{D}_l=\sigma_v^2\left(1-\frac{2}{\pi}\left(\frac{\gamma}{\gamma+1}\right)\right).
\end{eqnarray}
Note that for  $\gamma= 0$, we have $\bar{D}_l=\sigma_v^2$. On the other hand, in the high SNR regime, i.e., as $\gamma\rightarrow \infty$, we obtain $\bar{D}_l=\sigma_v^2(1-2/\pi)$. Both distortions can be argued to be asymptotically optimal. In fact, for zero SNR, even with an infinite-resolution front end, the MMSE estimate is given by $\hat{V}=0$, which yields $\bar{D}=\sigma_v^2$. Instead, for infinite SNR, the best mapping is given by the optimal binary quantizer, which yields $\bar{D}=\sigma_v^2(1-2/\pi)$ (see, e.g., \cite[Section 10.1]{Cover:book}).

\subsection{Digital Transmission for One-bit Front End}\label{Ch2_Sec:Digital Transmission}
Here we consider a conventional digital transmission scheme, which is based on quantizing and mapping the source to a discrete constellation for transmission over the channel. Accordingly, the source is quantized to one of the $M$ levels, each characterized by the interval $[v_{(l-1)},v_{(l)})$, $l=1,...,M$, where $v_{(M)}=\infty$, $v_{(0)}=-\infty$, and $v_{(l)}\geq v_{(l-1)}$ for all $l=1,...,M$. Each interval $[v_{(l-1)},v_{(l)})$ is mapped to the corresponding channel input $X=x_{(l)}$. We take the constellation of possible transmission points to be $\{X=A(2l-1-M), l=1,...,M\}$, for some parameter $A\geq 0$, such that the average power constraint is satisfied. Note that, when $M$ is even, this corresponds to the $M$-PAM modulation, while if $M$ is odd, the constellation includes the zero-power signal, i.e., $x_{\left(\frac{M+1}{2}\right)}=0$. The average transmission power can be written as $\mathbb{E}[X^2]=\sum_{l=1}^{M}x_{(l)}^2\cdot \mathrm{Pr}(X=x_{(l)})$, where $\mathrm{Pr}(X=x_{(l)})=\frac{1}{\sigma_v}\int_{v_{(l-1)}}^{v_{(l)}}\Phi\left(v/\sigma_v\right)dv$.

The average achievable distortion for $M$ levels of symmetric digital transmission, i.e., $v_{(l)}=-v_{(M-l)}$, can be easily obtained as
\begin{align}\label{Ch2_Eqn:12}
\bar{D}_{d,M}&=\sigma_v^2-\left(\frac{2}{\sigma_v}\sum\limits_{l=1}^{M}Q(S_l)\int\limits_{v_{(l-1)}}^{v_{(l)}}v\Phi\left(\frac{v}{\sigma_v}\right)dv\right)^2,
\end{align}
where $S_l^2=\frac{(2l-1-M)^2\gamma}{\sum_{l=1}^{M}(2l-1-M)^2\mathrm{Pr}(x_{(l)})}$.

As a special case, when $M=2$, setting the quantization threshold as $v_{(1)}=0$, we obtain BPSK transmission. The resulting achievable distortion can be computed from (\ref{Ch2_Eqn:12}) as
\begin{eqnarray}\label{Ch2_Eqn:13}
\bar{D}_{d,2}= \sigma_v^2\left(1-\frac{2}{\pi}\left(1-2Q\left(\sqrt{\gamma}\right)\right)^2\right).
\end{eqnarray}
We observe that, as for linear transmission, when $\gamma\rightarrow \infty$, we have $\bar{D}_{d,2}=\sigma_v^2(1-2/\pi)$, and when $\gamma\rightarrow0$, we have $\bar{D}_{d,2}=\sigma_v^2$. From (\ref{Ch2_Eqn:13}), one can check that the slope of the average distortion for BPSK transmission as $\gamma\rightarrow0$ is $-4\sigma_v^2/\pi^2$, whereas the slope for linear transmission, obtained from (\ref{Ch2_Eqn:8}), is $-2\sigma_v^2/\pi$. This means that linear transmission has a decline in the distortion in the low SNR regime; therefore, it outperforms BPSK transmission, in this regime.

As another example, for $M=3$, we set the quantization thresholds as $v_{(1)}=-c$ and $v_{(2)}=c$, so that $[-c,c]$ is the interval of source values for which the transmission symbol is $x_{(2)}=0$. The MSE distortion can be computed by solving the following optimization problem with line search:
\begin{eqnarray}
\bar{D}_{d,3}=\min\limits_{c\geq 0}~\sigma_v^2\left(1-\frac{2 e^{-\frac{c^2}{\sigma_v^2}}}{\pi}\cdot\kappa_\gamma(c)^2\right),
\end{eqnarray}
where $\kappa_\gamma(c)\triangleq 1-2Q\left(\sqrt{\frac{\gamma}{2Q\left(\frac{c}{\sigma_v}\right)}}\right)$.

\subsection{$K$-Level Front End}\label{Ch2_Sec:K_level_MSE}
In this section, we consider the system model in Figure \ref{Ch2_Figure:1} with a single observation, i.e., $N=1$, but with a $K$-level ADC front end as in (\ref{Ch2_Eqn:93}). As in the case of single one-bit observation, without loss of generality, we write the receiver mapping as
\begin{equation}\label{Ch2_Eqn:95}
g(Y)=\hat{V}=\hat{v}_{(j)},~\text{if}~ Y=y_{(j)},
\end{equation}
for $j=1,\ldots,K$. In the next proposition, we obtain a necessary optimality condition for the encoding and decoding functions $f$ and $g$.

\begin{prop}\label{Ch2_Result:2}
The optimal encoder and decoder mappings $f$ and $g$ for the $K$-level ADC front end in (\ref{Ch2_Eqn:93}) satisfy the necessary conditions
\begin{align}\nonumber
f(v)&=\frac{1}{2\sqrt{2\pi}\sigma_w\lambda}\sum\limits_{j=1}^{K}\left(\hat{v}_{(j)}\left(2v-\hat{v}_{(j)}\right)\vphantom{e^{-\frac{\left(z_{(j-1)}-f(v)\right)^2}{2\sigma_w^2}}}\right.\\\label{Ch2_Eqn:14}
&\left.\quad\cdot\left(e^{-\frac{\left(z_{(j-1)}-f(v)\right)^2}{2\sigma_w^2}}-e^{-\frac{\left(z_{(j)}-f(v)\right)^2}{2\sigma_w^2}}\right)\right),
\end{align}
and (\ref{Ch2_Eqn:95}) with
\begin{align}\label{Ch2_Eqn:66}
\hat{v}_{(j)}=\frac{\int v\Phi\left(\frac{v}{\sigma_v}\right)\left(Q\left(\frac{z_{(j-1)}-f(v)}{\sigma_w}\right)-Q\left(\frac{z_{(j)}-f(v)}{\sigma_w}\right)\right)dv}{\int \Phi\left(\frac{v}{\sigma_v}\right)\left(Q\left(\frac{z_{(j-1)}-f(v)}{\sigma_w}\right)-Q\left(\frac{z_{(j)}-f(v)}{\sigma_w}\right)\right)dv},
\end{align}
for $j=1,...,{K}$. Furthermore, the gradient of the Lagrangian function $L(f,g,\lambda)$ over $f$ for $g$ in (\ref{Ch2_Eqn:95}) is given as
\begin{align}\nonumber
  \nabla L&=2\lambda f(v)-\frac{1}{\sqrt{2\pi}\sigma_w}\sum\limits_{j=1}^{K}\left(\hat{v}_{(j)}\left(2v-\hat{v}_{(j)}\right)\vphantom{e^{-\frac{\left(z_{(j-1)}-f(v)\right)^2}{2\sigma_w^2}}}\right.\\\label{Ch2_Eqn:42}
  &\quad\cdot\left.\left(e^{-\frac{\left(z_{(j-1)}-f(v)\right)^2}{2\sigma_w^2}}-e^{-\frac{\left(z_{(j)}-f(v)\right)^2}{2\sigma_w^2}}\right)\right).
\end{align}
\end{prop}
\textit{Proof}: See Appendix \ref{Ch2_Appendix:2}.

As detailed later, the gradient in (\ref{Ch2_Eqn:42}), along with (\ref{Ch2_Eqn:66}), will be used in Section \ref{Ch2_Sec:Numerical results} to obtain numerically optimized encoders and decoders.

\section{Single Observation: Distortion Outage Probability}\label{Ch2_outage-single}
In this section, we study the optimal encoder and decoder under the DOP criterion defined in (\ref{Ch2_Eqn:2}) for the case of a single observation $(N=1)$. We first study the case of a one-bit front end in Section \ref{Ch2_Sec:Single_One_bit_DOP}, and then we extend the results to a $K$-level front end in Section \ref{Ch2_Sec:arbitrary ADC DOP}.

\subsection{Optimal Encoder and Decoder for a One-Bit Front End}\label{Ch2_Sec:Single_One_bit_DOP}
With no loss of generality, the decoder is given as in (\ref{Ch2_Eqn:5}) for some reconstruction points $(\hat{v}_{(1)},\hat{v}_{(2)})$. To proceed, we first focus on the optimization of the encoder mapping $f$ for a given decoder in (\ref{Ch2_Eqn:5}). We then tackle the problem of minimizing the DOP over the reconstruction points $(\hat{v}_{(1)},\hat{v}_{(2)})$.

\begin{figure}
\begin{centering}
\includegraphics[scale=0.22]{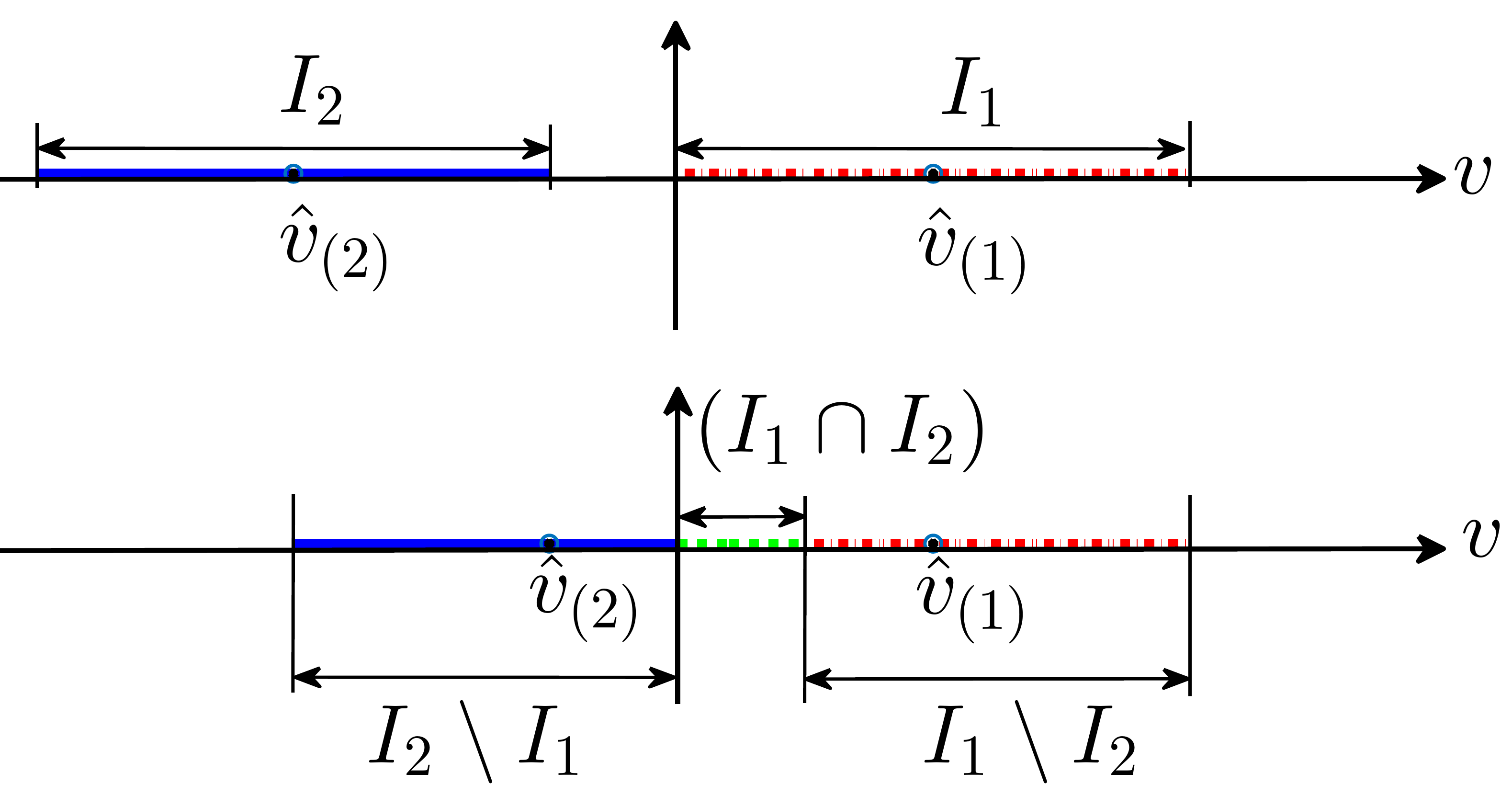}
\caption{Illustration of the intervals $I_1$, $I_2$ and $(I_1 \cap I_2)$ that characterize the optimal encoder for the DOP criterion, for two different cases depending on the $(\hat{v}_{(1)},~\hat{v}_{(2)})$ values.}\label{Ch2_Figure:2}
\par\end{centering}
\vspace{0mm}
\end{figure}

To elaborate, we define the intervals
\begin{align}\label{Ch2_Eqn:16}
I_j\triangleq\{v:(v-\hat{v}_{(j)})^2 < D\},
\end{align}
for $j=1,2$, which are depicted in Figure \ref{Ch2_Figure:2}. Each interval $I_j$, corresponds to the set of source values that are within the allowed distortion $D$ of the reconstruction point $\hat{v}_{(j)}$. The following claims hold: (\textit{i}) For all source outputs $v$ in the set $(I_1\cup I_2)^{C}=\{v:\min_{j=1,2}(v-\hat{v}_{(j)})^2 > D\}$, outage occurs (superscript $C$ denotes the complement set). We refer to this event as \textit{source outage}. (\textit{ii}) For all source values in the interval $I_1 \cap I_2$, either of the reconstruction points yield a distortion no more than the target value $D$. Therefore, regardless of which of the two reconstruction levels, $\hat{v}_{(1)}$ and $\hat{v}_{(2)})$, is selected by the receiver, no outage occurs. From observations (\textit{i}) and (\textit{ii}), it easily follows that, for all source values $v$ inside the intervals $(I_1\cup I_2)^{C}$ and $(I_1\cap I_2)$, the optimal mapping is $f(v)=0$, since, for both intervals, the occurrence of an outage event is independent of the transmitted signal.

From the discussion above, we only need to specify the optimal mapping for the intervals $I_1\backslash I_2$ and $I_2\backslash I_1$. This should be done by accounting not only for the \textit{source outage} event mentioned above, but also for the \textit{channel outage} events. In particular, the distortion outage probability $\epsilon(D)$ can be written as
\begin{align}\nonumber
\epsilon(D)&=\mathrm{Pr}\left(V\in (I_1\cup I_2)^{C}\right)\\\nonumber
&\quad+\mathrm{Pr}\left(V\in (I_1\setminus I_2),\hat{V}=\hat{v}_{(2)}\right)\\\label{Ch2_Eqn:17}
&\quad+\mathrm{Pr}\left(V\in (I_2\setminus I_1),\hat{V}=\hat{v}_{(1)}\right),
\end{align}
where the first term accounts for the source outage event, while the second and third terms are the probabilities of outage due to channel transmission errors. For instance, the second term is the probability that the decoder selects $\hat{V}=\hat{v}_{(2)}$ while $V$ is in the interval $I_1\setminus I_2$ (see Figure \ref{Ch2_Figure:2}). The next proposition characterizes the optimal encoder mapping.

\begin{prop}\label{Ch2_Result:3}
Given a target distortion $D$, and arbitrary reconstruction points $\hat{v}_{(1)}$ and $\hat{v}_{(2)}$, the optimal mapping $f$ for the problem (\ref{Ch2_Eqn:3}) is given by
\begin{align}\label{Ch2_Eqn:18}
f(v)&=\left\{\begin{array}{ll}
0&v\in  (I_1\cup I_2)^{C}\cup (I_1\cap I_2),\\
-u&v\in (I_2\backslash I_1),\\
u&v\in (I_1\backslash I_2),
\end{array}\right.
\end{align}
where $u$ is the unique solution of
\begin{align}\label{Ch2_Eqn:19}
ue^{\frac{u^2}{2\sigma_w^2}}=\frac{1}{2\sqrt{2\pi}\sigma_w\lambda}.
\end{align}
\end{prop}
\textit{Proof}: See Appendix \ref{Ch2_Appendix:3}.

We note here that, for given $\lambda\geq 0$, the optimal $u$ is independent of the values of $\hat{v}_{(1)}$ and $\hat{v}_{(2)}$. Examples of optimal encoders will be provided in Section \ref{Ch2_Sec:Numerical results}. In the next proposition, we turn to the optimization of the reconstruction levels $(\hat{v}_{(1)},\hat{v}_{(2)})$.

\begin{prop}\label{Ch2_Result:4}
The optimal reconstruction points $(\hat{v}_{(1)},\hat{v}_{(2)})$, are given by
\begin{subequations}\label{Ch2_Eqn:20}
\begin{align}\label{Ch2_Eqn:21}
  \hat{v}_{(1)}&=\sqrt{D}-a^{*},\\
  \hat{v}_{(2)}&=-\hat{v}_{(1)},
\end{align}
\end{subequations}
where $a^{*}$ is obtained from
\begin{align}\nonumber
a^{*}&=\!\!\!\!\!\stackunder[5pt]{arg min}{$a\in[0,\sqrt{D}]$}\!\!2Q\left(\frac{2\sqrt{D}-a}{\sigma_v}\right)\!+\!2\left(Q\left(\frac{u}{\sigma_w}\right)\!+\!\lambda u^2\right)\\\label{Ch2_Eqn:22}
&\quad\quad\quad\quad\quad\quad\quad\cdot\left(Q\left(\frac{a}{\sigma_v}\right)-Q\left(\frac{2\sqrt{D}-a}{\sigma_v}\right)\right),
\end{align}
where $u$ is obtained by solving (\ref{Ch2_Eqn:19}).
\end{prop}
\textit{Proof}: See Appendix \ref{Ch2_Appendix:4}.

To summarize, the optimal encoder and decoder are obtained as follows. First, given the Lagrange multiplier $\lambda \geq 0$, the value of $u$ is obtained by solving (\ref{Ch2_Eqn:19}). Then the decoder's reconstruction points $(\hat{v}_{(1)},\hat{v}_{(2)})$ are computed from (\ref{Ch2_Eqn:20})-(\ref{Ch2_Eqn:22}). Finally, the optimal encoder mapping is given by (\ref{Ch2_Eqn:18}). The next remark elaborates on the optimal encoder and decoder in two asymptotic SNR regimes.
\begin{rem}\label{Ch2_Remark:2}
If $\lambda$ is large, i.e., in the low-SNR regime, from (\ref{Ch2_Eqn:19}) we have $u\approx 0$. Also from (\ref{Ch2_Eqn:19}) it can be verified that $\lambda u^2=\frac{e^{-\frac{u^2}{\sigma_w^2}}}{8\pi\sigma_w^2 \lambda}\approx 0$. Hence, from (\ref{Ch2_Eqn:22}) we obtain
\begin{align}\label{Ch2_Eqn:23}
a^{*} \approx \stackunder[5pt]{arg min}{$a\in[0,\sqrt{D}]$}Q\left(\frac{a}{\sigma_v}\right)+Q\left(\frac{2\sqrt{D}-a}{\sigma_v}\right),
\end{align}
yielding $\hat{v}_{(1)}=\hat{v}_{(2)}=0$, that is, $I_1=I_2$ and $\epsilon(D)=2Q\left(\frac{\sqrt{D}}{\sigma_v}\right)$. On the other hand, for small values of $\lambda$, corresponding to the high-SNR regime, the variable $u$ becomes large. Also, from (\ref{Ch2_Eqn:19}) we have the equality $\lambda =1\big/\left(2\sqrt{2\pi}\sigma_w u e^{\frac{u^2}{2\sigma_w^2}}\right)$ and Hence the approximations $\lambda u^2=u/\left(2\sqrt{2\pi}\sigma_w e^{\frac{u^2}{2\sigma_w^2}}\right)\approx0$, and
\begin{align}\label{Ch2_Eqn:24}
a^{*}\approx\stackunder[5pt]{arg min}{$a\in[0,\sqrt{D}]$}2Q\left(\frac{2\sqrt{D}-a}{\sigma_v}\right),
\end{align}
which yield distinct intervals $I_1$ and $I_2$ with $\hat{v}_{(1)}=-\hat{v}_{(2)}=\sqrt{D}$, and  $\epsilon(D)=2Q\left(\frac{2\sqrt{D}}{\sigma_v}\right)$.
\end{rem}

\subsection{$K$-Level Front End}\label{Ch2_Sec:arbitrary ADC DOP}
Here we turn our attention to the case of $K$-level front end under the DOP criterion. With no loss of optimality, the decoder is given as in (\ref{Ch2_Eqn:95}) for some reconstruction levels $\{\hat{v}_{(1)},\ldots,\hat{v}_{(K)}\}$ to be optimized. For a subset $\mathcal{V}\subset\{\hat{v}_{(1)},...,\hat{v}_{(K)}\}$, let $I_{\mathcal{V}}$ be the set of source outputs $v$ for which the quadratic distance between $v$ and the reconstruction points in set ${\mathcal{V}}$ is less than $D$, while the quadratic distance with respect to the reconstruction points in $\mathcal{V}^C$ is larger than $D$. This set is defined as
\begin{equation}\label{Ch2_Eqn:94}
I_{\mathcal{V}}= \bigcap_{\hat{v}_{(j)}\in \mathcal{V}}I_{j}\setminus \bigcup_{\hat{v}_{(j)}\in \mathcal{V}^C} I_{j},~|\mathcal{V}|=1,...,K,
\end{equation}
where $I_{j}$ is given as
\begin{align}
I_{j}\triangleq\left\{v:|v-\hat{v}_{(j)}|^2<D\right\}.
\end{align}
We also define the set $I_{\emptyset}$ corresponding to ${\mathcal{V}}=\emptyset$ as
\begin{equation}
I_{\emptyset}\triangleq\left(\bigcup_{{\mathcal{V}}:|{\mathcal{V}}|\neq 0} I_{\mathcal{V}}\right)^{C},
\end{equation}
that is, the set of values $v\in \mathbb{R}$ that do not have a reconstruction value $\hat{v}_{(j)}$ within a distance $\sqrt{D}$. Note that the sets $\left\{I_{\mathcal{V}}:~\mathcal{V}\subset\{\hat{v}_{(1)},...,\hat{v}_{(K)}\}\right\}$ form a partition of the whole real line.

As for the single one-bit front end (\ref{Ch2_Eqn:17}), the DOP depends on both source and channel outage events. Note that the source outage occurs if $V\in I_{\emptyset}$, whereas the channel outage occurs when $V\notin I_{\emptyset}$ and $V\notin I_{\{\hat{v}_{(1)},\ldots,\hat{v}_{(K)}\}}$, with no outage occurring when $V\in I_{\{\hat{v}_{(1)},\ldots,\hat{v}_{(K)}\}}$. In the following proposition, we present necessary optimality conditions for the encoder and decoder mappings.

\begin{prop}\label{Ch2_Result:5}
For a $K$-level ADC front end, the optimal encoder and decoder mappings $f$ and $g$ satisfy the necessary conditions
\begin{eqnarray}\label{Ch2_Eqn:65}
f(v)&=\frac{1}{2\lambda}G(f,v),
\end{eqnarray}
where $G(f,v)$ is defined as
\begin{align}\nonumber
&G(f,v)\triangleq\\\label{Ch2_Eqn:81}
&\left\{\begin{array}{ll}
0 &v\in I_{\mathcal{V}}: |{\mathcal{V}}|=0,K,\\
        \sum\limits_{j:~\hat{v}_{(j)}\in {\mathcal{V}}}\frac{\xi(j,f(v))}{\sqrt{2\pi}\sigma_w} &v\in I_{\mathcal{V}}:~|{\mathcal{V}}|=1,\ldots,K-1,
\end{array}\right.
\end{align}
where $\xi(j,f(v))\triangleq e^{-\frac{\left(z_{(j-1)}-f(v)\right)^2}{2\sigma_w^2}}-e^{-\frac{\left(z_{(j)}-f(v)\right)^2}{2\sigma_w^2}}$,
and (\ref{Ch2_Eqn:95}) with
\begin{align}\nonumber
\!\!\!\hat{v}_{(j)}&=\arg\max_{t}~\int\limits_{t-\sqrt{D}}^{t+\sqrt{D}}\Phi\left(\frac{v}{\sigma_v}\right)\\\label{Ch2_Eqn:96}
&\quad~\quad\quad\cdot\left(\!Q\!\left(\!\frac{z_{(j-1)}\!-\!f(v)}{\sigma_w}\!\right)\!-\!Q\!\left(\!\frac{z_{(j)}\!-\!f(v)}{\sigma_w}\!\right)\!\right)dv.
\end{align}
Furthermore, the gradient of the Lagrangian function $L(f,g,\lambda)$ over $f$ for $g$ in (\ref{Ch2_Eqn:95}) is given as
\begin{eqnarray}\label{Ch2_Eqn:10}
\nabla L=2\lambda f(v)-\frac{1}{2} G(f,v).
\end{eqnarray}
\end{prop}
\textit{Proof}: See Appendix \ref{Ch2_Appendix:5}.

As for Proposition \ref{Ch2_Result:2}, in Section \ref{Ch2_Sec:Numerical results}, we will use (\ref{Ch2_Eqn:96}) and (\ref{Ch2_Eqn:10}) to obtain numerically optimized encoders and decoders, that satisfy the necessary optimality conditions.

\section{Multiple Observations}\label{Ch2_Sec:Multiple Observation}
In this section, we study the more general case in which the receiver has $N>1$ noisy one-bit quantized observations, $Y^N$, of the transmitted source sample. For both the MSE distortion and the DOP criteria, without loss of generality, we write the receiver mapping as
\begin{equation}\label{Ch2_Eqn:98}
g(Y^N)=\hat{V}=\hat{v}_{(j)},~\text{if}~~ Y^N=b_{j}^N,
\end{equation}
where $b_{j}^N$ is the $N$-length binary representation of the number $2^N-j$ for $j=1,\ldots,2^N$. Note that there are $2^N$ reconstruction levels $\hat{v}_{(j)}$, each of which corresponds to a different configuration of the received signal $Y^N\in\{0,1\}^N$. We will denote the $k$-th element of the vector $b_j^N$ by $b_j^N(k)$.

\subsection{MSE Criterion}\label{Ch2_Sec:Multi_obs_MSE}
Recalling that the optimal decoding function under the MSE criterion is the MMSE estimator, the optimal decoder satisfies $\hat{v}_{(j)}=\mathbb{E}[V|Y^N=b_{j}^N]$. In the next proposition, we provide necessary conditions for the optimal encoder and decoder mappings along with an expression for the gradient of the Lagrangian in (\ref{Ch2_Eqn:4}) over the encoder mapping $f$.

\begin{prop}\label{Ch2_Result:6}
Given a front end with $N$ one-bit ADCs, the optimal encoder and decoder mappings $f$ and $g$, respectively, satisfy the necessary conditions
\begin{align}\label{Ch2_Eqn:15}
 f(v)&=\frac{1}{2\lambda}\sum\limits_{j=1}^{2^N}\Theta \left(N,f(v),j\right)\hat{v}_{(j)} \left(2v-\hat{v}_{(j)}\right),
\end{align}
and (\ref{Ch2_Eqn:98}) with
\begin{align}\label{Ch2_Eqn:67}
\hat{v}_{(j)}&\triangleq\frac{\int v\Phi\left(\frac{v}{\sigma_v}\right)\prod\limits_{i=1}^{N}Q\left(\frac{(-1)^{b_{j}^N(i)+1}f(v)}{\sigma_{w_i}}\right)dv}{\int  \Phi\left(\frac{v}{\sigma_v}\right)\prod\limits_{i=1}^{N}Q\left(\frac{(-1)^{b_{j}^N(i)+1}f(v)}{\sigma_{w_i}}\right)dv},
\end{align}
where $\Theta \left(N,f(v),j\right)$ is defined as
\begin{align}\nonumber
\!\!\!\!\Theta \left(N,f(v),j\right)&\triangleq\sum\limits_{k=1}^{N}\left(\frac{(-1)^{b_{j}^N(k)}e^{-\frac{f(v)^2}{2\sigma_{w_k}^2}} }{\sqrt{2\pi}\sigma_{w_k}}\right.\\
&\quad\quad\left.\cdot\prod\limits_{l=1,l\neq k}^{N}Q\left(\frac{(-1)^{b_{j}^N(l)+1}f(v)}{\sigma_{w_l}}\right)\right).
\end{align}
Furthermore, the gradient of the Lagrangian function $L(f,g,\lambda)$ over $f$ for $g$ in (\ref{Ch2_Eqn:98}) is given as
\begin{align}\label{Ch2_Eqn:37}
  \nabla L=2\lambda f(v)-\sum\limits_{j=1}^{2^N}\hat{v}_{(j)}\Theta \left(N,f(v),j\right)\left(2v-\hat{v}_{(j)}\right).
\end{align}
\end{prop}
\textit{Proof}: See Appendix \ref{Ch2_Appendix:6}.

\begin{rem}\label{rem1}
In order to obtain numerically optimized encoder (NOE) mappings in Section \ref{Ch2_Sec:Numerical results}, we will adopt a gradient-descent algorithm. This method updates the current iterate $f^{(i)}(v)$ as
\begin{align}\label{Ch2_Eqn:62}
  f^{(i+1)}(v)=f^{(i)}(v)-\mu \nabla L,
\end{align}
where $i$ is the iteration index and $\mu>0$ is the step size. $\nabla L$ is the derivative of the Lagrangian (\ref{Ch2_Eqn:3}) with respect to the mapping  $f$ at $f^{(i)}$, which can be found in (\ref{Ch2_Eqn:42}) and (\ref{Ch2_Eqn:37}) for $K$-level ADC and multiple one-bit ADCs, respectively. The algorithm is initialized with the linear mapping specified in (\ref{Ch2_Eqn:9}). It is noted that the algorithm is not guaranteed to converge to a global optimal solution.
\end{rem}

\subsection{DOP Criterion}\label{Ch2_Sec:Multiple Observation_DOP}
In this subsection, we consider the DOP criterion. The analysis follows the same steps as in Section \ref{Ch2_Sec:arbitrary ADC DOP}. In particular, we define the different intervals $I_{\mathcal{V}}$, where ${\mathcal{V}}$ is a subset of $\{\hat{v}_{(1)},\ldots,\hat{v}_{\left(2^N\right)}\}$ as in (\ref{Ch2_Eqn:94}). Based on this definition, in the next proposition, we derive necessary optimality conditions for the encoder and decoder mappings.

\begin{prop}\label{Ch2_Result:7}
Given a front end with $N$ one-bit ADCs, the optimal encoder and decoder mappings $f$ and $g$ satisfy the necessary conditions
\begin{eqnarray}\label{Ch2_Eqn:25}
f(v)=\frac{1}{2\lambda}\tilde{G}(f,v),
\end{eqnarray}
where $\tilde{G}(f,v)$ is defined as
\begin{align}\nonumber
&\tilde{G}(f,v)\triangleq\\\label{Ch2_Eqn:83}
&\left\{\begin{array}{ll}
\!\!\!0 & v\!\in \! I_{\mathcal{V}}\!:\!|{\mathcal{V}}|\!=\!0,2^N,\\
\!\!\!\sum\limits_{k:\hat{v}_{(k)}\in {\mathcal{V}}}\!\left(\!\sum\limits_{i=1}^{N}\frac{(-1)^{b_{k}^N(i)}e^{-\frac{f(v)^2}{2\sigma_{w_i}^2}}}{\sqrt{2\pi}\sigma_{w_i}}\right.&v\!\in\! I_{\mathcal{V}}\!: \!|{\mathcal{V}}|\!=\!1,...,2^N\!\!-\!1,\\
\quad\quad\cdot\left.\prod\limits_{\substack{l=1\\l\neq i}}^{N}\!Q\!\left(\!\frac{(-1)^{b_{k}^N(l)+1}f(v)}{\sigma_{w_l}}\!\right)\!\!\right)&
     \end{array}\right.
\end{align}
and (\ref{Ch2_Eqn:98}) with
\begin{align}\label{Ch2_Eqn:97}
\hat{v}_{(j)}&=\!\arg\max_{t}\!\int\limits_{t-\sqrt{D}}^{t+\sqrt{D}}\!\Phi\!\left(\!\frac{v}{\sigma_v}\!\right)
\prod\limits_{i=1}^{N}Q\left(\frac{(-1)^{b_{j}^N(i)+1}f(v)}{\sigma_{w_i}}\right)dv,
\end{align}
for $j=1,\ldots,2^N$. Furthermore, the gradient of the Lagrangian function $L(f,g,\lambda)$ over $f$ for $g$ in (\ref{Ch2_Eqn:98}) is given as
\begin{eqnarray}\label{Ch2_Eqn:74}
\nabla L=2\lambda f(v)-\frac{1}{2} \tilde{G}(f,v).
\end{eqnarray}
\end{prop}
\textit{Proof}: See Appendix \ref{Ch2_Appendix:7}.

\begin{rem}\label{Ch2_Remark:3}
From the optimal decoders in (\ref{Ch2_Eqn:67}) and (\ref{Ch2_Eqn:97}), under the MSE and DOP criteria, respectively, it can be easily verified that, with equal noise variances for the $N$ observations, i.e., $\sigma_{w_i}^2=\sigma_w^2$ for $i=1,...,N$, the reconstruction points $\hat{v}_{(j)}$ corresponding to $b_j^N$ vectors with the same number of ones are equal. As a consequence, there are only $N+1$ effective  reconstruction points instead of $2^N$.
\end{rem}

\begin{rem}\label{rem2}
In Section \ref{Ch2_Sec:Numerical results}, we obtain NOE mappings using the gradient-descent method as follows:
\begin{enumerate}
  \item Initialize the set of reconstruction points $\{\hat{v}_{(j)}\},~j=1,\ldots,2^B$;
  \item Optimize the encoder mapping corresponding to the decoder (\ref{Ch2_Eqn:95}) and (\ref{Ch2_Eqn:98}) for $K$-level ADC and multiple one-bit ADCs, respectively, for the given $\{\hat{v}_{(j)}\},~j=1,\ldots,2^B$, using the gradient-descent algorithm (\ref{Ch2_Eqn:62}), with $\nabla L$ defined as in (\ref{Ch2_Eqn:10}) for $K>2$, and as in (\ref{Ch2_Eqn:74}) for $N>1$;
  \item Find the optimal reconstruction points corresponding to the obtained encoder mapping using (\ref{Ch2_Eqn:96}) for $K>2$, and (\ref{Ch2_Eqn:97}) for $N>1$;
  \item If convergence is not obtained, go back to step 2.
\end{enumerate}
\end{rem}

\begin{figure}
\begin{centering}
\includegraphics[scale=0.47]{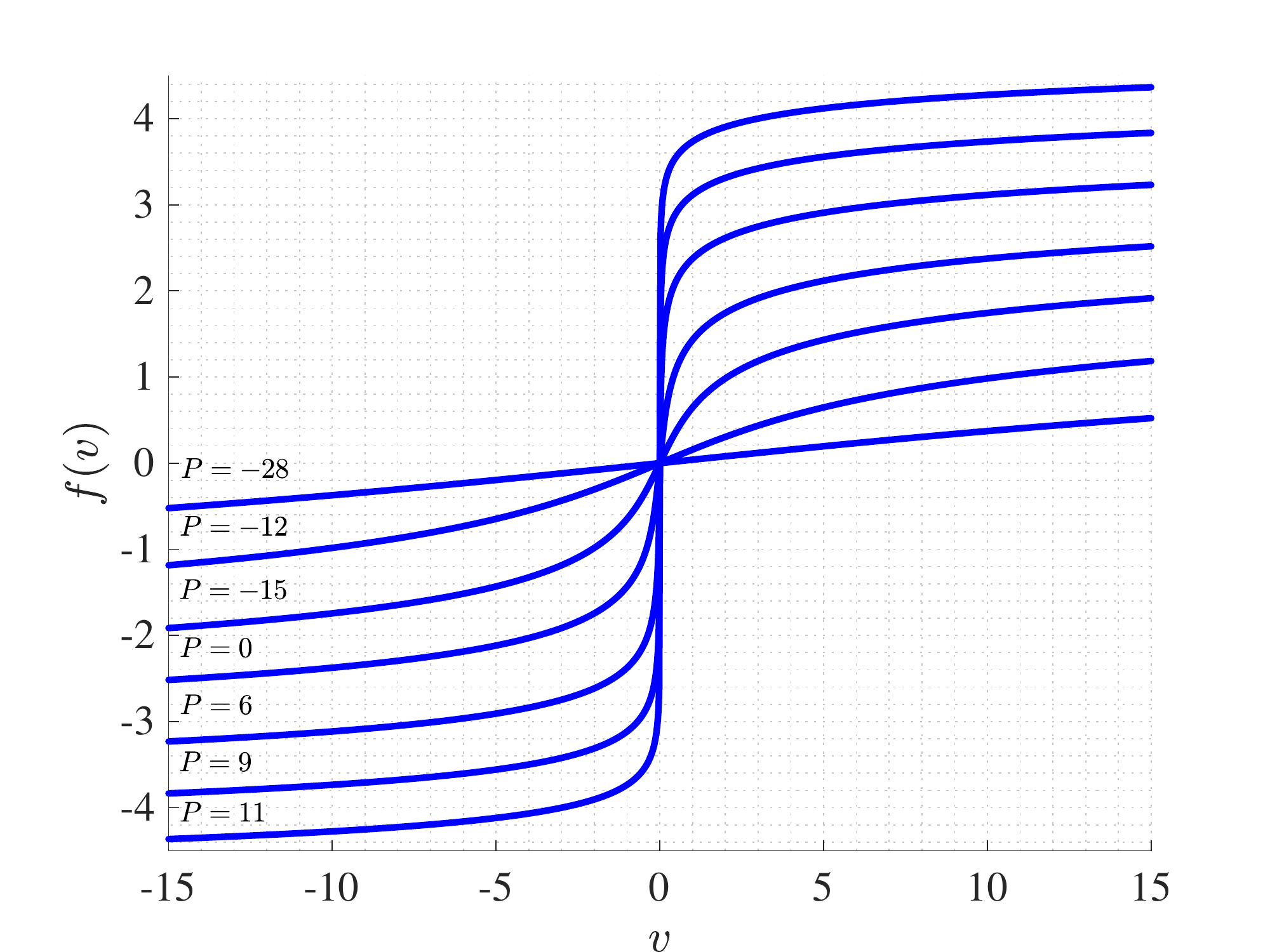}
\caption{Optimal encoder mappings for the MSE distortion criterion under different power constraints $P$ (dB) ($\sigma_w^2=\sigma_v^2=1$).}\label{Ch2_Figure:6}
\par\end{centering}
\vspace{0mm}
\end{figure}

\section{Numerical Results}\label{Ch2_Sec:Numerical results}
In this section, we provide some illustrations of the results derived above by means of numerical examples. We consider the MSE distortion and the DOP in separate subsections. Throughout this section, we set $\sigma_v^2=\sigma_w^2=1$, so that the SNR in (\ref{Ch2_Eqn:75}) is proportional to the power constraint $P$.

\subsection{MSE Criterion}
We start by considering the MSE distortion in a single measurement system $(N=1)$ with a one-bit ADC front end $(K=2)$. We then investigate alternative front ends with more levels ($K>2$) or more observations ($N>1$) and present a performance comparison of different front end architectures.

For the case $N=1$ and $K=2$, Figure \ref{Ch2_Figure:6} shows the optimal mappings obtained from Proposition \ref{Ch2_Result:1} for different values of $P$. The value of the Lagrange multiplier $\lambda$ in (\ref{Ch2_Eqn:6}) is obtained by means of bisection so as to satisfy the power constraint $\mathbb{E}[f(V)^2]=P$. Figure \ref{Ch2_Figure:6} shows that the optimal mapping has a linear behaviour around the origin, and that it saturates as the absolute value of the source sample increases. As discussed in Section \ref{Ch2_Sec:Optimality Conditions}, for low SNR, the saturation occurs only for very large absolute values of the source; and therefore, the numerically obtained solution for (\ref{Ch2_Eqn:6}) tends to a linear function, while, for high SNR, the numerically obtained solution tends to a step function, resembling digital transmission.

\begin{figure}
\begin{centering}
\includegraphics[scale=0.47]{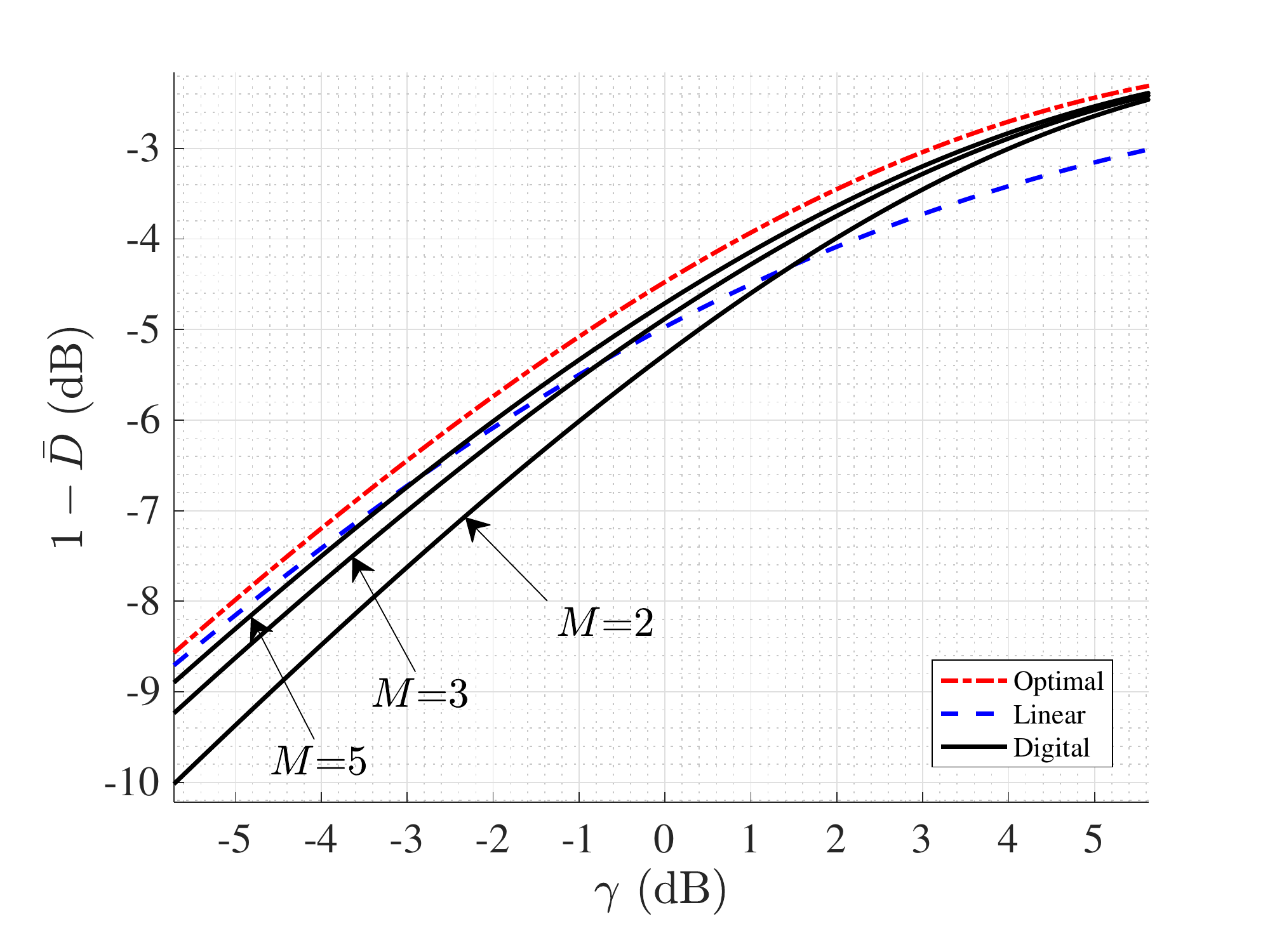}
\caption{Complement of the MSE $(1-\bar{D})$ (dB) versus the SNR $(\gamma)$ (dB) for the optimal encoder obtained in Proposition \ref{Ch2_Result:1} compared with the linear and digital schemes studied in Sections \ref{Ch2_Sec:Optimality Conditions}, \ref{Ch2_Sec:Uncoded transmission} and \ref{Ch2_Sec:Digital Transmission} ($\sigma_w^2=\sigma_v^2=1$).}\label{Ch2_Figure:7}
\par\end{centering}
\vspace{0mm}
\end{figure}

The conclusions above are corroborated by Figure \ref{Ch2_Figure:7}, in which the MSE distortion of the optimal, linear and digital transmission schemes are plotted versus the SNR $\gamma$. For clarity of illustration, we plot the complementary MSE distortion $1-\bar{D}$, where we note that $\bar{D}=\sigma_v^2=1$ is achievable by setting $\hat{V}=0$ irrespective of the received signal. The figure confirms that linear transmission approaches optimality at low SNR, whereas, for high SNR, digital schemes outperform linear transmission. It is also seen that, for digital transmission, increasing the number of constellation points generally improves the performance, although, in the high SNR regime, binary transmission is sufficient to achieve the minimum MSE.

We now investigate the MSE performance with the $K$-level ADC $(N=1,~K>2)$ front end studied in Section \ref{Ch2_Sec:K_level_MSE} and with the multiple one-bit ADCs front end $(N>1,~K=2)$ considered in Section \ref{Ch2_Sec:Multi_obs_MSE}. For these architectures, for which Proposition \ref{Ch2_Result:2} and Proposition \ref{Ch2_Result:6} provide respective necessary optimality conditions, we resort to a gradient-descent approach, as described in Remark \ref{rem1}.

\begin{figure}[h!]
\begin{centering}
\begin{subfigure}{.5\textwidth}
  \centering
  \includegraphics[scale=0.45]{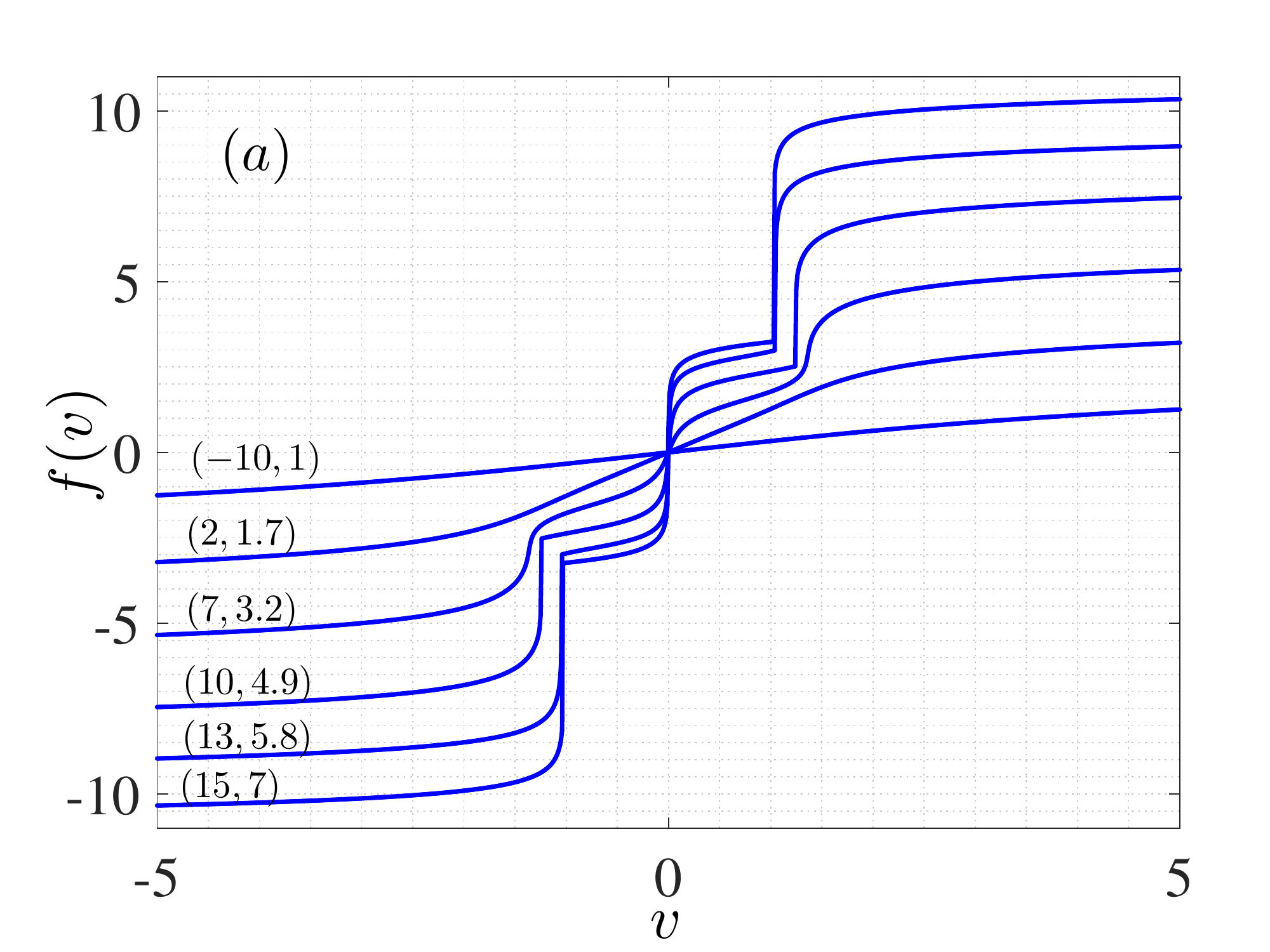}
\end{subfigure}
\begin{subfigure}{.5\textwidth}
  \centering
  \includegraphics[scale=0.45]{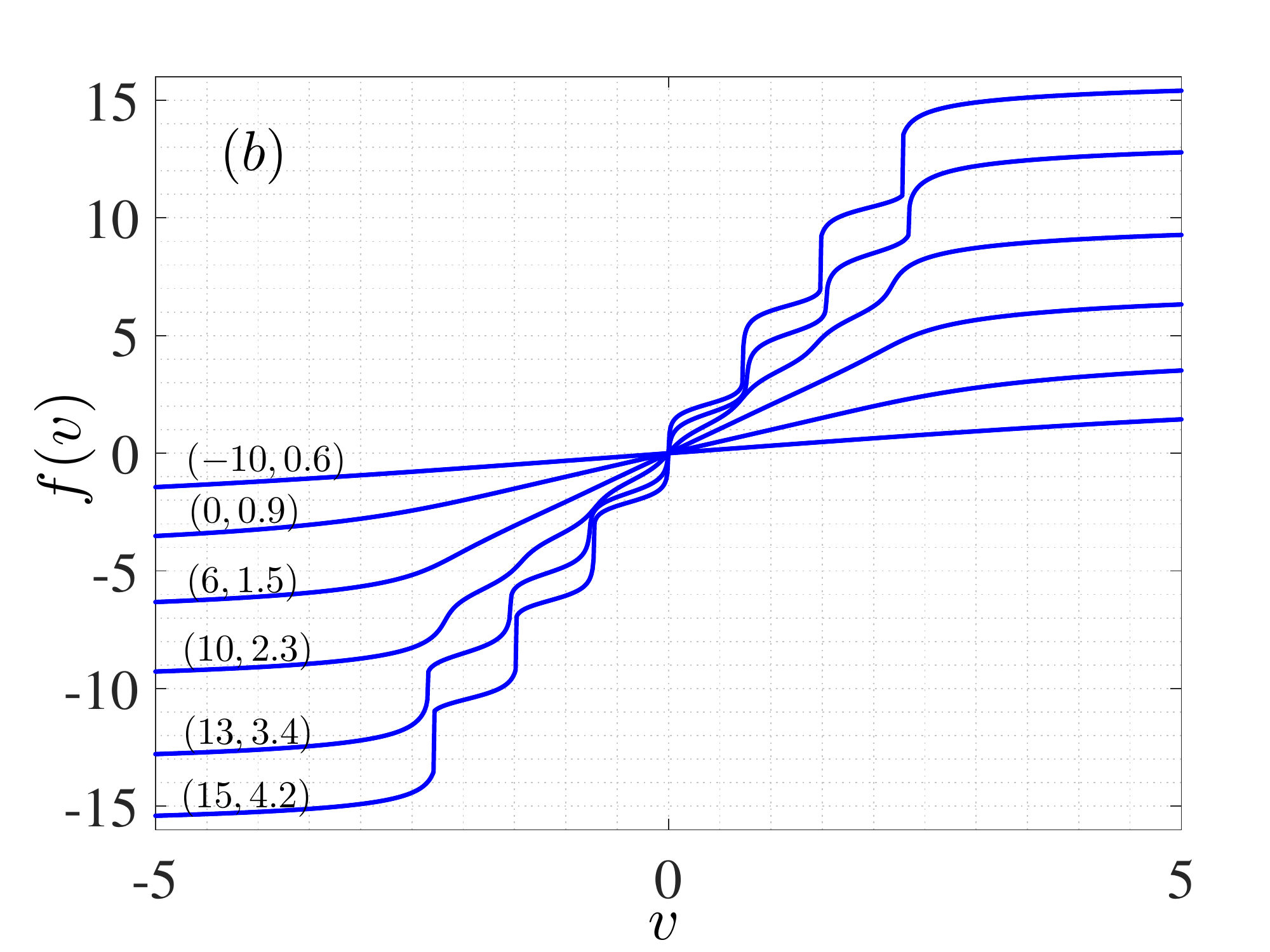}
\end{subfigure}
\caption{NOE mappings for different power constraints (dB) and under the MSE distortion criterion ($\sigma_w^2=\sigma_v^2=1$): $(a)$ $K=4$; $(b)$ $K=8$. The curves are labelled by the pair $(P,d)$, where $P$ is the average power constraint (dB) and $d$ is the quantization step size of the ADC front end.}\label{Ch2_Figure:8}
\end{centering}
\end{figure}

\begin{figure}[h!]
\begin{centering}
\begin{subfigure}{.5\textwidth}
  \centering
  \includegraphics[scale=0.45]{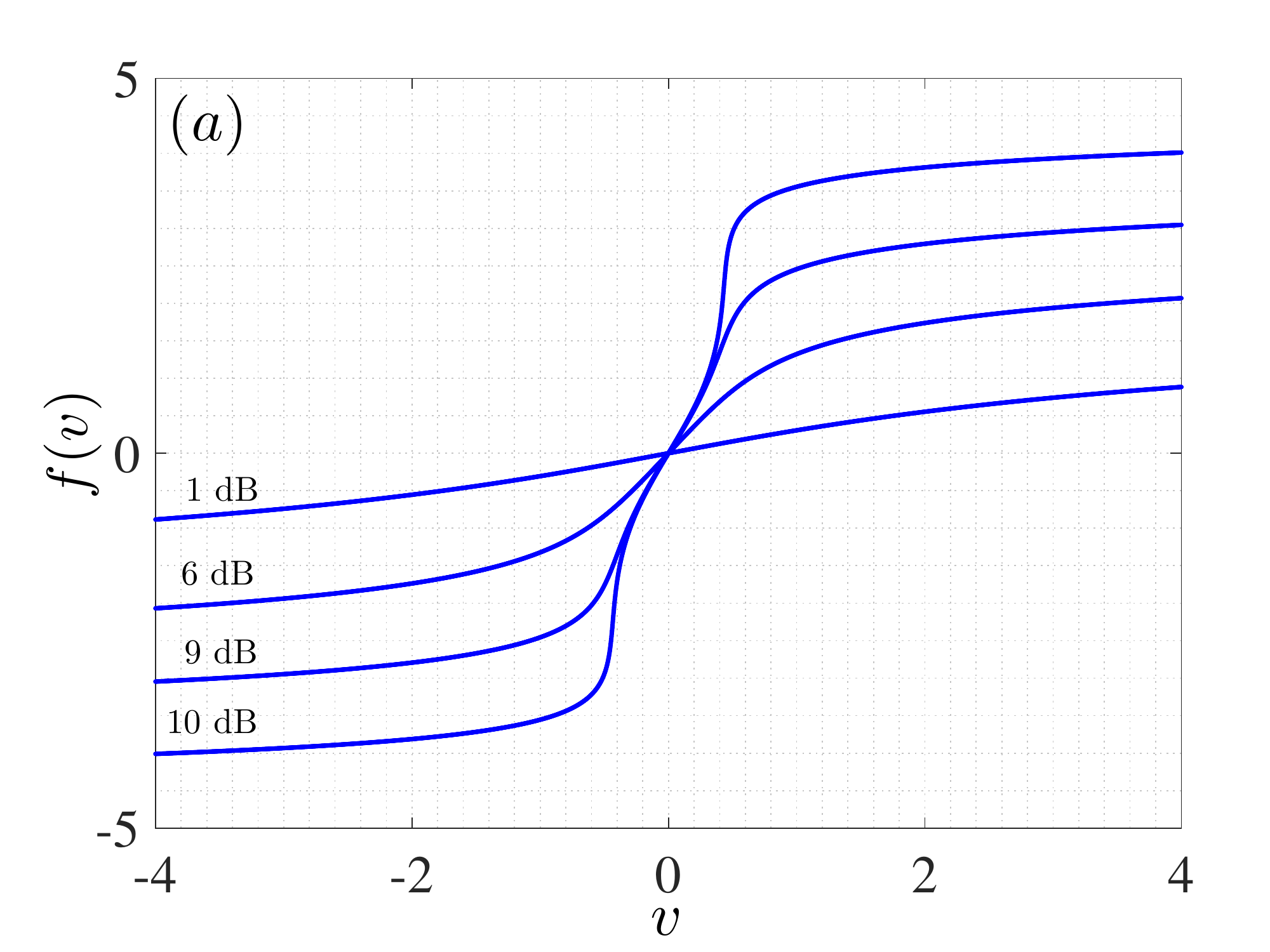}
\end{subfigure}
\begin{subfigure}{.5\textwidth}
  \centering
  \includegraphics[scale=0.37]{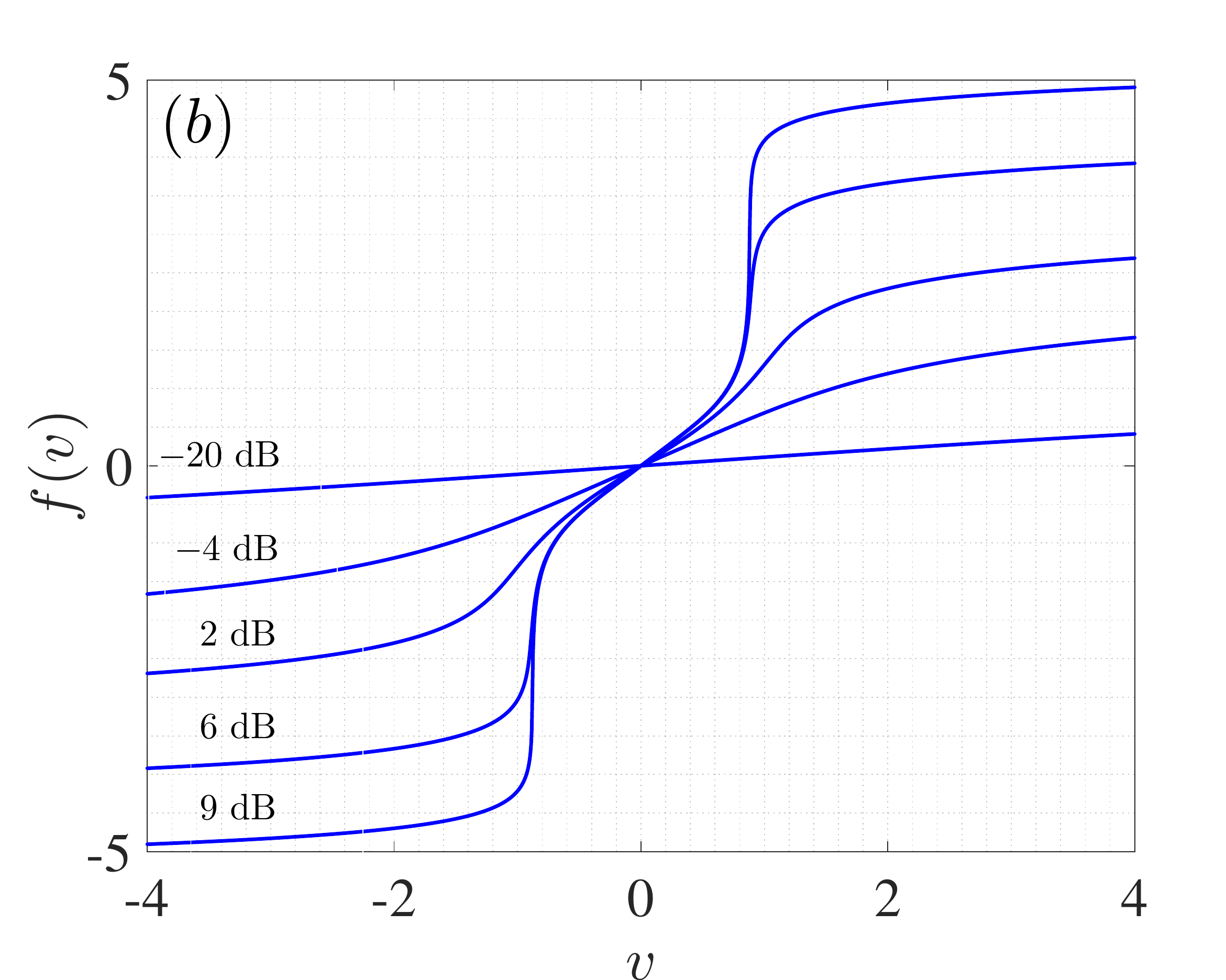}
\end{subfigure}
\caption{NOE mappings for different power constraints (dB) and under the MSE distortion criterion $\sigma_w^2=\sigma_v^2=1$: $(a)$ $N=2,~K=2$; $(b)$, $N=7,~K=2$. The curves are labelled by their corresponding average power constraint (dB).}\label{Ch2_Figure:15}
\end{centering}
\end{figure}

We now first illustrate the NOE mappings for the two types of front end architectures and then provide a performance comparison. In Figure \ref{Ch2_Figure:8}, the obtained NOE mappings are illustrated for a $K=4$ and a $K=8$ level ADC front end, respectively, under the MSE distortion criterion. The decision thresholds for $K=4$ and $K=8$ are chosen as $[\infty,-d,0,d,\infty]$ and as $[-\infty,-3d,-2d,-d,0,d,2d,3d,\infty]$, respectively, and the parameter $d$ is optimized by means of a line search. From the results obtained in Figure  \ref{Ch2_Figure:8}, it is noticed that, in the low SNR regime, the NOE mappings for both $K=4$ and $K=8$ approach linear transmission as discussed in Section \ref{Ch2_Sec:Optimality Conditions}. Instead, as the SNR $\gamma$ increases, they resemble digital mappings with $K$ constellation points.

Considering now the alternative architecture with $N$ ADCs each with $K=2$ quantization levels, Figure \ref{Ch2_Figure:15} illustrates the NOE mappings for $N=2$ and $N=7$ under different power constraints. It is again observed that, when the SNR is low, the NOE mappings approach linear transmission. With increasing SNR the NOE mapping exhibits a saturation behaviour, although, as opposed to the $N=1$ case in Figure \ref{Ch2_Figure:6}, the linear behaviour around the origin does not disappear completely.

\begin{figure}
\begin{centering}
\includegraphics[scale=0.47]{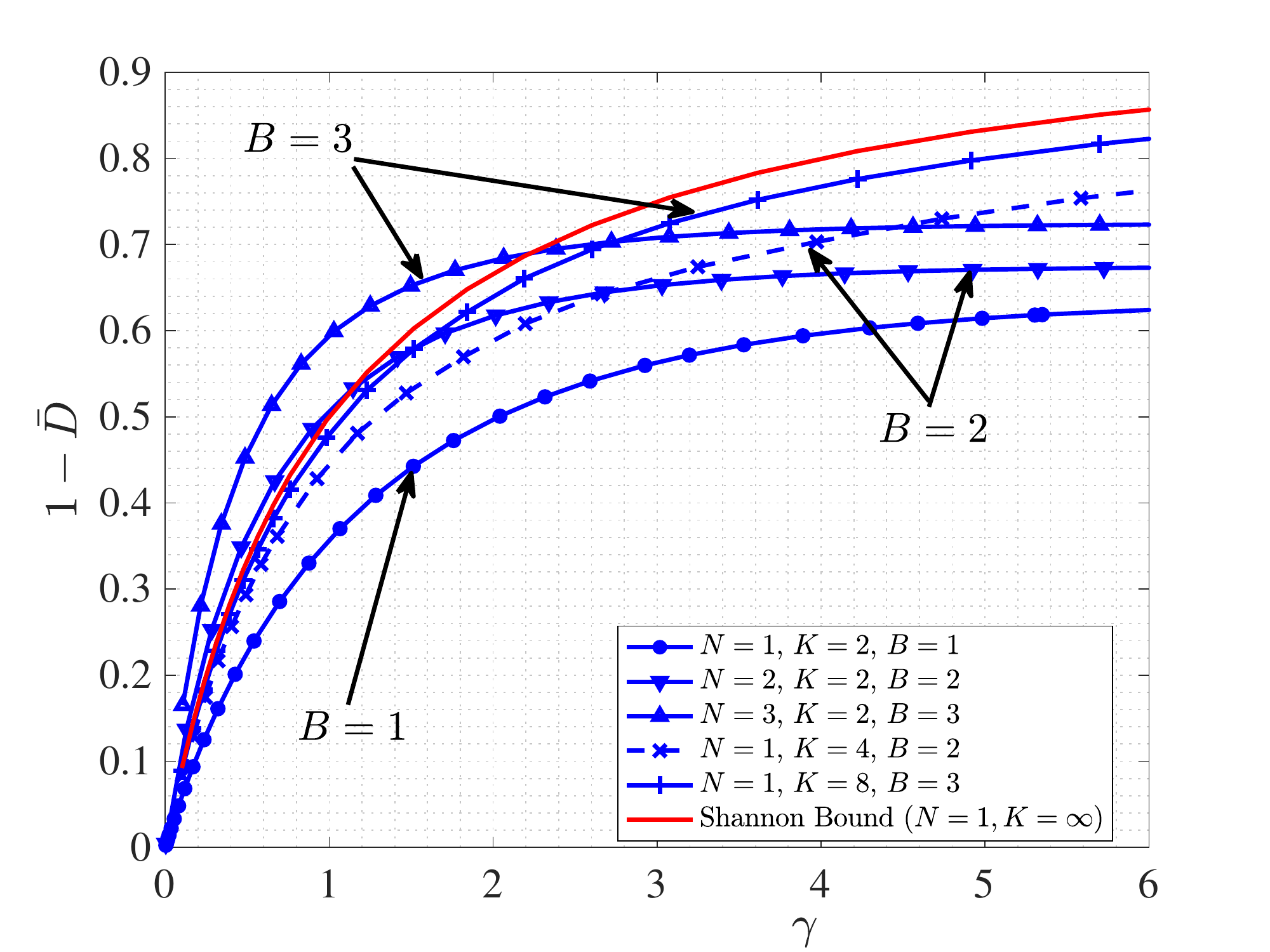}
\caption{Complement of the MSE $(1-\bar{D})$ vs. SNR $(\gamma)$ (linear scale) for different ADC architectures with $B$-bit outputs ($\sigma_v^2=1$). The variance of the AWGN in all the scenarios and for all the observations is one $(\sigma_w^2=1)$.}\label{Ch2_Figure:12}
\par\end{centering}
\vspace{0mm}
\end{figure}

We now compare the two front end architectures, under the constraint that they both provide $B$ output bits per received sample. The first architecture uses all bits to quantize a single observation, i.e., $N=1$ and $K=2^B$, while the second one outputs $B$ one-bit measurements, i.e., $N=B,~K=2$. In Figure \ref{Ch2_Figure:12}, the achievable MSE distortion is shown for the two architectures for $B=1,~2,~3$, along with the Shannon bound \cite[Equation 21]{Goblick:IT:65}, which corresponds to the theoretical performance limit for $N=1$ and $K=\infty$. It is seen that, for the same number of bits per sample, $B$, as the SNR $(\gamma)$ increases, the $2^B$-level ADC architecture outperforms the receiver with $B$ one-bit ADCs, whereas, the opposite is true for low SNR. This shows that, for high SNR, it is more beneficial to invest additional output bits in improving the ADC resolution. In contrast, for low SNR, it is preferable to increase the number of observations in order to improve the effective SNR by collecting independent measurements of the transmitted signal. We also note that, we found this conclusion to hold when the threshold values of the $N$ one-bit ADCs are allowed to be distinct, and optimized. Similar observations have also been made when the mappings are restricted to be linear (not shown here).

\subsection{DOP Criterion}
\begin{figure}[h!]
\begin{centering}
\begin{subfigure}{.5\textwidth}
  \centering
  \includegraphics[scale=0.45]{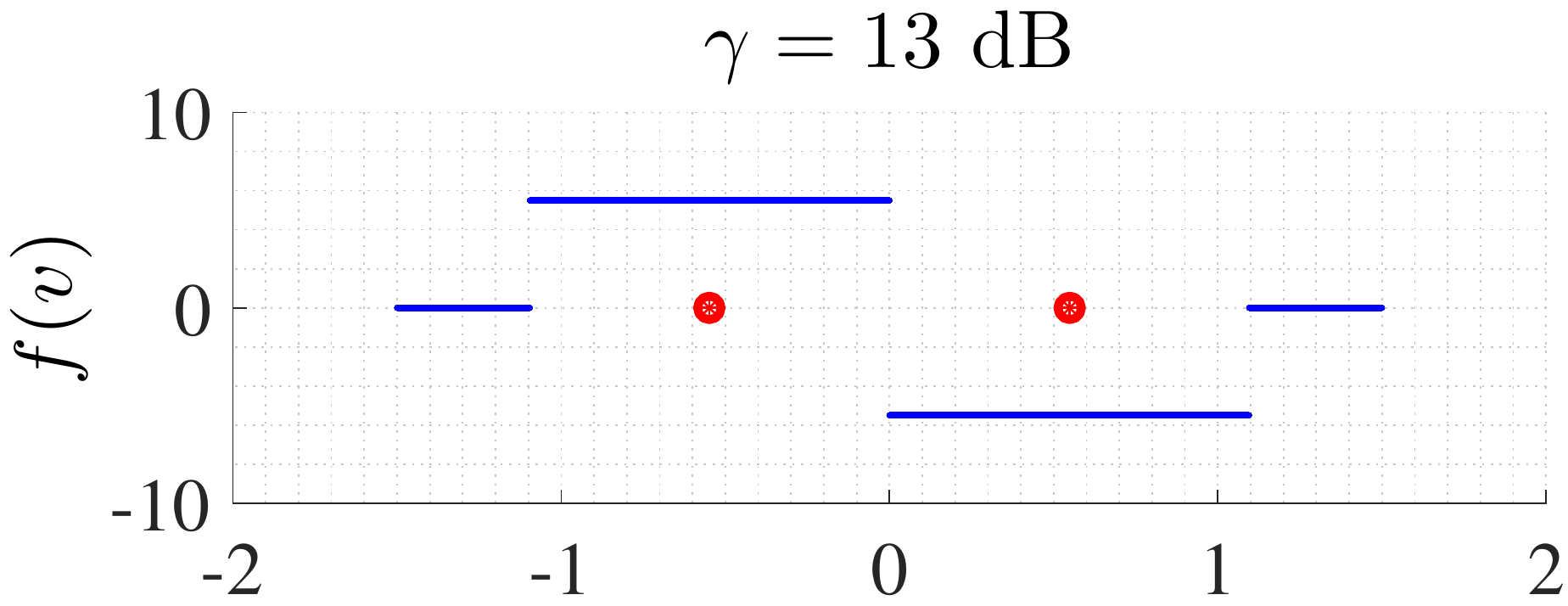}
\end{subfigure}
\begin{subfigure}{.5\textwidth}
  \centering
  \includegraphics[scale=0.45]{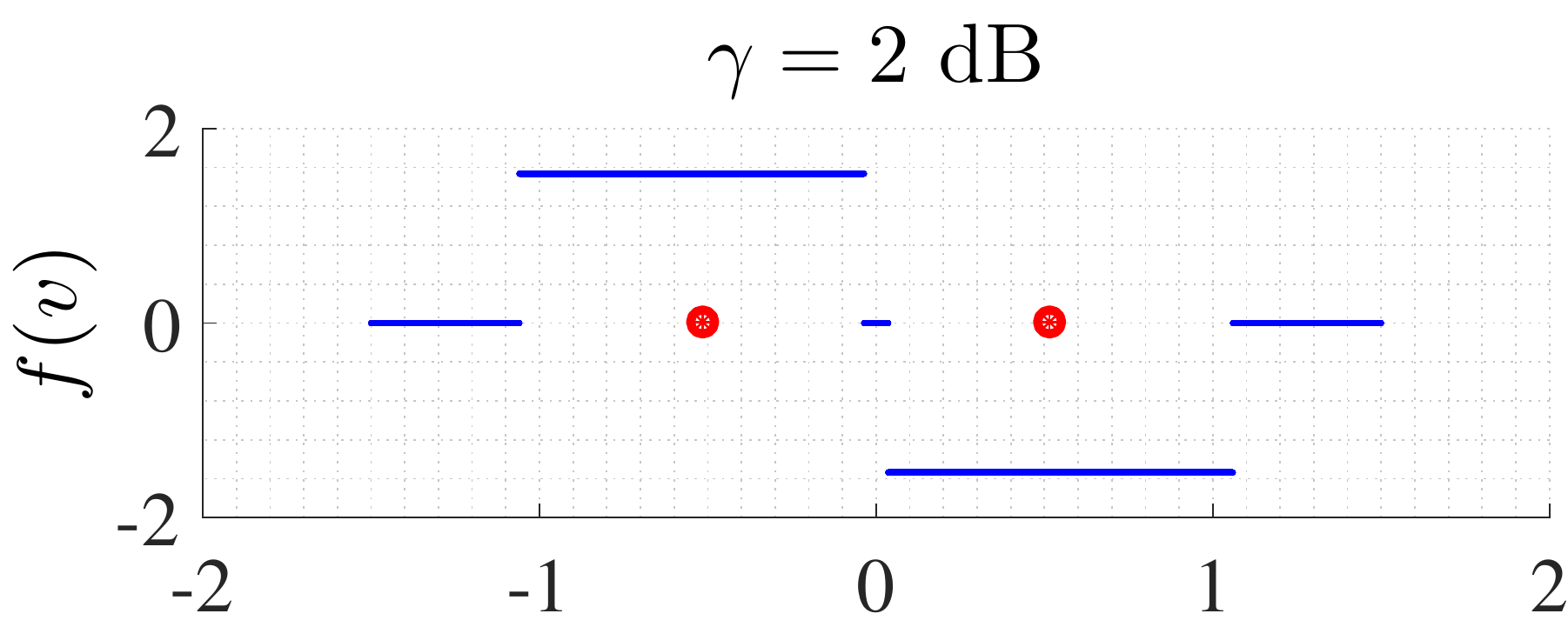}
\end{subfigure}
\begin{subfigure}{.5\textwidth}
  \centering
  \includegraphics[scale=0.45]{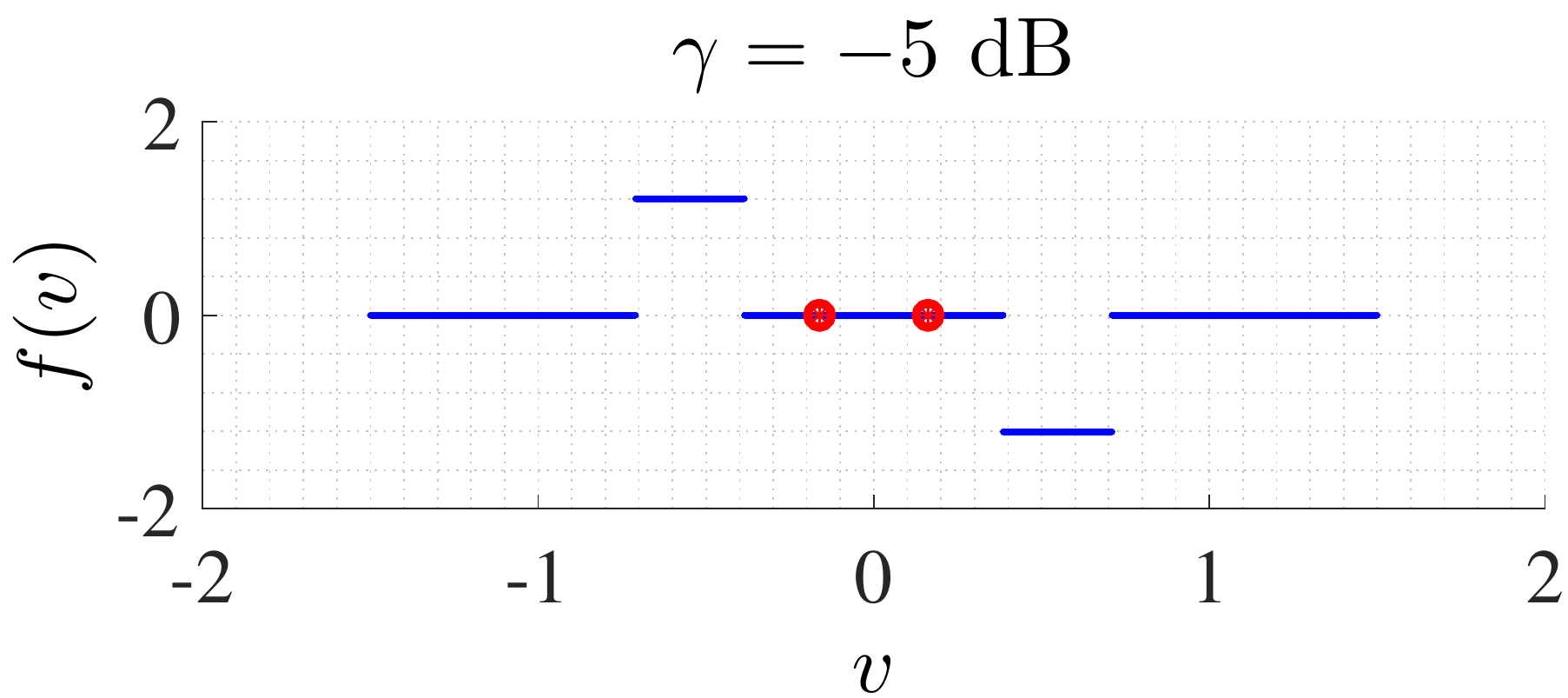}
\end{subfigure}
\caption{Optimal encoder mappings for $N=1,~K=2$ under the DOP criterion, for different values of the SNR $(\gamma)$. The dots are the reconstruction points $\hat{v}_{(1)}$ and $\hat{v}_{(2)}$ ($\sigma_w^2=\sigma_v^2=1$ and $D=0.3$).}\label{Ch2_Figure:13}
\end{centering}
\end{figure}

Here, we study the optimal mapping and performance under the DOP criterion. We start by considering the case of a one-bit ADC front end with a single observation, i.e., $N=1$, $K=2$. The optimal mapping along with the corresponding optimal reconstruction points $(\hat{v}_{(1)},\hat{v}_{(2)})$ for three different values of the power constraint $P$ are shown in Figure \ref{Ch2_Figure:13}, as obtained in Propositions \ref{Ch2_Result:3} and \ref{Ch2_Result:4}. It is seen that, as the SNR decreases, the optimal reconstruction points, $\hat{v}_{(1)}$ and $\hat{v}_{(2)}$, tend to zero, while, for high SNR, they tend to $\hat{v}_{(1)}=-\hat{v}_{(2)}=\sqrt{D}$, as per Remark \ref{Ch2_Remark:2}. This observation can be explained as follows. For low SNR, the DOP is dominated by the probability of channel outage, i.e., by the last two terms in (\ref{Ch2_Eqn:17}), which are zero for $\hat{v}_{(1)}=\hat{v}_{(2)}=0$. In contrast, for high SNR, the optimal solution aims at minimizing the probability of source outage events, i.e., the first term in (\ref{Ch2_Eqn:17}), which requires $\hat{v}_{(1)}=-\hat{v}_{(2)}=\sqrt{D}$.

\begin{figure}
\begin{centering}
\includegraphics[scale=0.47]{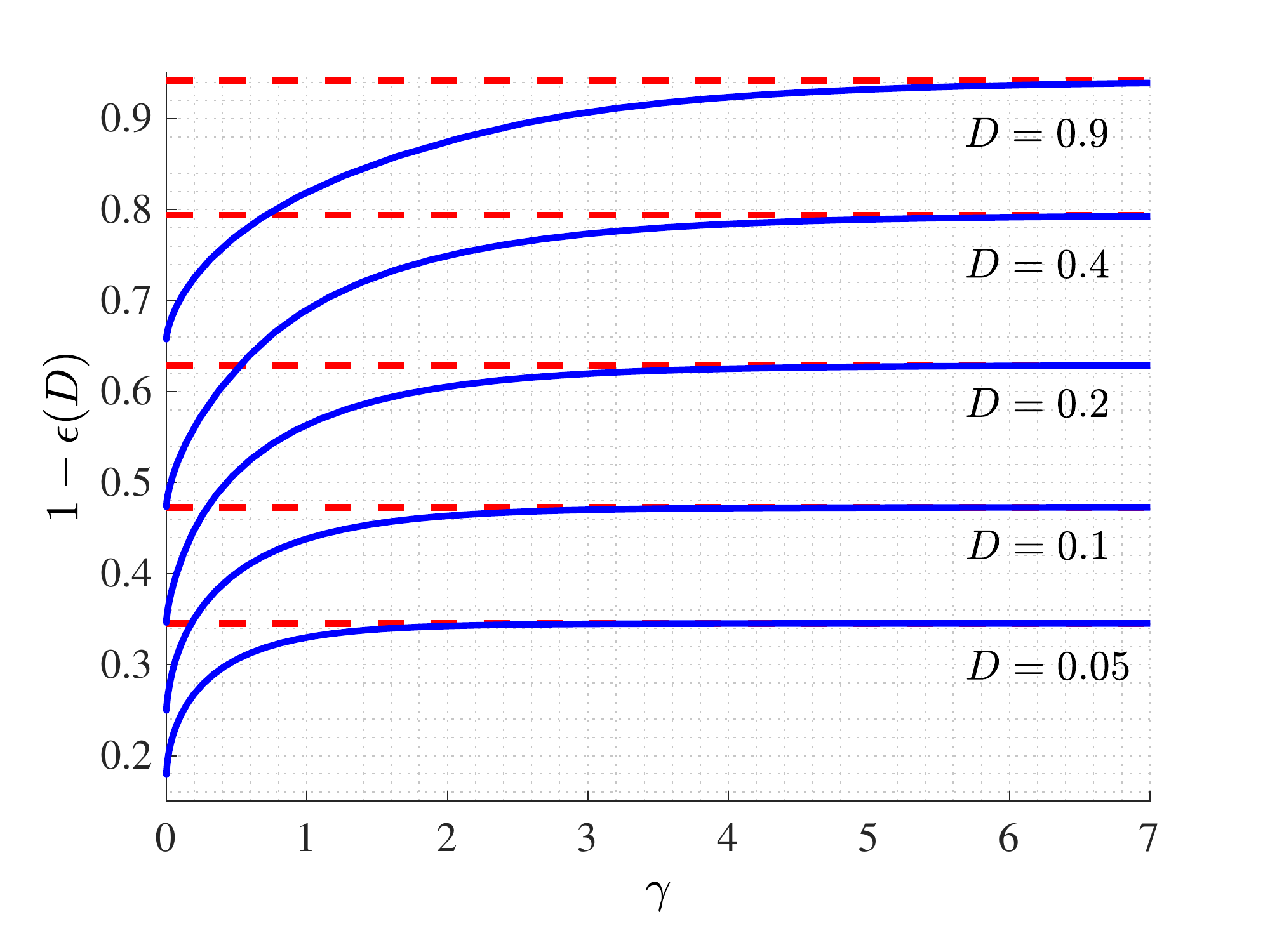}
\caption{Complement of the DOP $(1-\epsilon(D))$ versus SNR $(\gamma)$ in linear scale. Dashed lines represent the DOP in the high SNR regime for the corresponding distortion target ($\sigma_w^2=\sigma_v^2=1$).}\label{Ch2_Figure:14}
\par\end{centering}
\vspace{0mm}
\end{figure}

Continuing the analysis of the $N=1,~K=2$ case, in Figure \ref{Ch2_Figure:14}, the complement of the DOP, $1-\epsilon(D)$, is plotted with respect to the SNR for different values of $D$. As SNR decreases, based on the discussion above and Remark \ref{Ch2_Remark:2}, DOP tends to the first term in (\ref{Ch2_Eqn:17}) when $\hat{v}_{(1)}=\hat{v}_{(2)}=0$, which can be computed as $Q(\sqrt{D}/\sigma_v)$. Furthermore, as the SNR increases, DOP tends to the first term in (\ref{Ch2_Eqn:17}) but with  $\hat{v}_{(1)}=-\hat{v}_{(2)}=\sqrt{D}$, resulting in $\epsilon(D)=Q(2\sqrt{D}/\sigma_v)$, indicated by the dashed lines in Figure \ref{Ch2_Figure:14}.

In order to derive NOE mapping under the DOP criterion for $K>2$ or $N>1$, we apply iterative gradient-descent based algorithm based on the results obtained in Propositions \ref{Ch2_Result:5} and \ref{Ch2_Result:7} as described in Remark \ref{rem2}.

For low SNR, based on the results in Section \ref{Ch2_Sec:Single_One_bit_DOP} (see also Figure \ref{Ch2_Figure:13}) the reconstruction points are close to zero; and hence, we can initialize the algorithm with all-zero values when $\lambda$ is very large. Therefore, we first set a large value for $\lambda$ and consider all-zero vector as the initial mapping. Then, we consider successively smaller values of $\lambda$, i.e., increase the SNR. We use the reconstruction points obtained for the previous value of the Lagrange multiplier $\lambda$ to initialize the algorithm for the current value of $\lambda$. This approach is known as noise channel relaxation (NCR) \cite{Gadkari_Rose_1999}.

\begin{figure}
\begin{centering}
\includegraphics[scale=0.47]{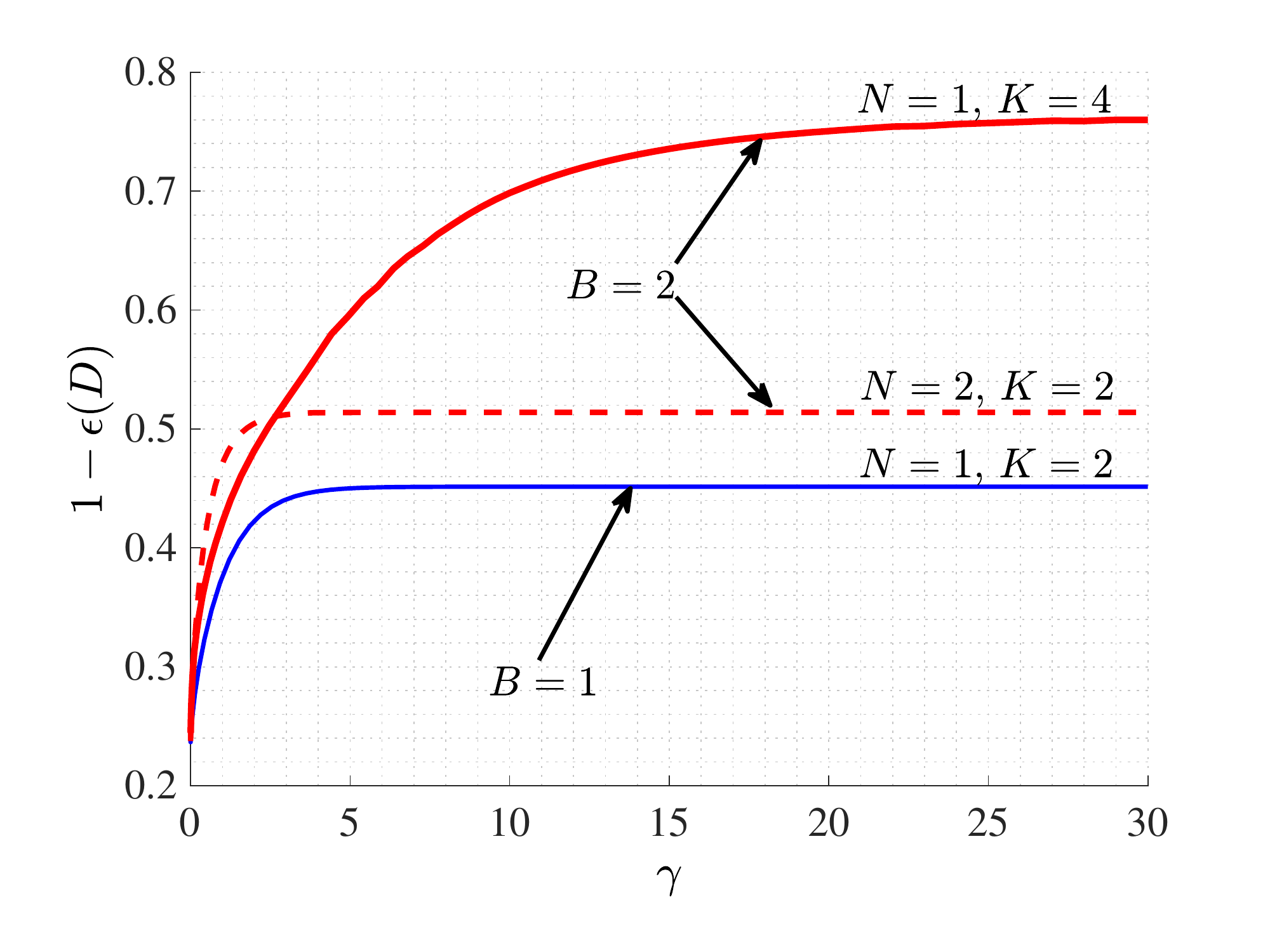}
\caption{Complement of the DOP $(1-\epsilon(D))$ versus SNR $(\gamma)$ in linear scale for different ADC architectures with $B$-bit outputs ($\sigma_w^2=\sigma_v^2=1$).}\label{Ch2_Figure:18}
\par\end{centering}
\vspace{0mm}
\end{figure}

In Figure \ref{Ch2_Figure:18}, the complement of the DOP, $1-\epsilon(D)$, is shown for two architectures with $B=2$ output bits, namely one observation with 4-level ADC $(N=1$, $K=4)$, and two observations with one-bit ADCs $(N=2$, $K=2)$, as well as for the architecture with a one-bit ADC $(N=1,~K=2)$, for a target distortion of $D=0.09$. In accordance with the discussion above, all architectures have the same performance for low SNR, namely $\epsilon(D)=2Q(\sqrt{D}/\sigma_v)=2Q(\sqrt{0.09})$. Furthermore, in a manner similar to the discussion on Figure \ref{Ch2_Figure:12} for the MSE criterion, in the low SNR regime, it is beneficial to increase the number of observations, whereas at high SNR, it is preferable to increase the ADC resolution. In this regard, we note that, in the high SNR the optimal $K=4$ levels are selected to minimize the probability of source outage, that is, the probability  $\textrm{Pr}(V\in I_{\emptyset})$, yielding the reconstruction points $\hat{V}=\{-3\sqrt{D},-\sqrt{D},\sqrt{D},3\sqrt{D}\}$ and a DOP of $2Q(4\sqrt{D}/\sigma_v)=2Q(1.2)=0.2301$. Instead, for $N=2$ and $K=2$, the three effective reconstruction points (see Remark \ref{Ch2_Remark:3}) at high SNR tend to $\hat{V}=\{-2\sqrt{D},0,2\sqrt{D}\}$ and the minimum DOP is lower bounded by $Q(3\sqrt{D}/\sigma_v)=2Q(.9)=0.3681$.

\section{Conclusions and Future Work}\label{Ch2_Sec:Conclusion}
We have considered the zero-delay transmission of a single sample of a Gaussian source over a quantized SIMO AWGN channel. We first studied a system with a one-bit ADC front end at the receiver under two distinct performance criteria, namely the mean square error (MSE) distortion and the distortion outage probability (DOP). For the MSE distortion, we have shown that the optimal encoder mapping is odd, and that, in the low SNR regime, linear transmission approaches the optimal performance, whereas digital transmission becomes optimal in the high SNR regime. For the DOP criterion, we have obtained the optimal structure of the encoder and the decoder, demonstrating that the optimal encoder function is symmetric and piecewise constant. For both the MSE distortion and the DOP criteria, we also derived necessary optimality conditions for the encoder and decoder mappings for a $K$-level ADC front end and for multiple one-bit ADC observations.

Among open problems that are left for future research, we mention here the joint optimization of the quantization intervals of the ADC front end with the encoding and decoding functions. Another interesting problem is the zero-delay joint source-channel coding over fading channels with finite-resolution ADCs.



\section{Appendices}\label{Ch2_Sec:app}
\appendix 
\section{Preliminaries: Calculus of Variations}
In the proofs of the propositions reported above, we leverage the standard method in variational calculus to obtain necessary optimality conditions \cite[Section 7]{Luenberger_1969}. The following lemma presents the key result that will be used throughout the following appendices. For the sake of brevity, we drop the arguments of functions and functionals where no confusion can arise. We also use the notation $F^f$ and $F^{f^{'}}$ to denote the partial derivatives of a functional $F$ with respect to functions $f$ and $f^{'}$, respectively.

\begin{lem}\label{Ch2_Result:8}
Let $G_i,~i=1,\ldots,n$, be continuous functionals of $(f$, $f^{'}$, $u)$ and have continuous partial derivatives with respect to $(f$, $f^{'})$. Also, let $F$ be a continuous functional of $(f$, $f^{'}$, $\fa,\ldots, \fn, u)$, where $\oi$ is a functional of $G_i$ given as
\begin{equation}\label{Ch2_Eqn:82}
\oi =\int\limits_{t_1}^{t_2} G_i(f(t),f^{'}(t),t)dt,~~i=1,\ldots,n.
\end{equation}
Let $F$ has continuous partial derivatives with respect to $(f,f^{'},\fa,\ldots, \fn)$. Consider the following minimization problem
\begin{eqnarray}\label{Ch2_Eqn:72}
\begin{aligned}
& \underset{f}{\text{minimize}}
& & L(f)\triangleq\int\limits_{t_1}^{t_2} F(f(t),f^{'}(t),\fa,\ldots, \fn ,t)dt.
\end{aligned}
\end{eqnarray}
Define the functional derivative $\nabla L$ as
\begin{align}\nonumber
\nabla L\triangleq& F^{f}(f(u),f^{'}(u),\fa,\ldots, \fn ,u)\\\nonumber
&-\frac{d}{du}F^{f^{'}}(f(u),f^{'}(u),\fa,\ldots, \fn ,u)\\\nonumber
&+\!\sum\limits_{i=1}^{n}\!\bigg(\!\left(G_i^{f}(f(u),f^{'}(u),u)\!-\!\frac{d}{du}G_i^{f^{'}}(f(u),f^{'}(u),u)\right)\\
&\quad\quad\quad\cdot\int F^{\oi }(f(t),f^{'}(t),\fa,\ldots, \fn ,t)dt\bigg),
\end{align}
where $F^{r_i}$ is the derivative of the functional $F$ with respect to $r_i$. Similarly, $G_i^{f},G_{i}^{f^{'}}$ are derivatives of the functional $G_{i}$ with respect to $f,f^{'}$, respectively. A necessary condition for a function $f$ to be a solution of the problem (\ref{Ch2_Eqn:72}) is
\begin{align}\label{Ch2_Eqn:84}
  \nabla L=0,~\forall u\in[t_1,t_2].
\end{align}
\end{lem}
\textit{Proof}: Following the conventional approach in the calculus of variations, we perturb the function $f(t)$ by an arbitrary function $\eta(t)$, which vanishes on the boundary points $t_1$ and $t_2$ \cite{Luenberger_1969}. Let $\delta_f L\triangleq \frac{dL(f+\alpha\eta)}{d\alpha}\Big|_{\alpha=0}$ be the Gateaux derivative of the functional $L$ with respect to the parameter $\alpha$. We have
\begin{align}\nonumber
&\!\!\!\!\!\delta_f  L=\\\label{Ch2_Eqn:106}
&\!\!\!\!\!\frac{d}{d\alpha}\!\int\limits_{t_1}^{t_2}\! F(f(t)+\alpha \eta(t),f^{'}(t)+\alpha {\eta}^{'}(t),\fa^\alpha ,\ldots, \fn^\alpha,t)dt\Bigg|_{\alpha=0}\!\!,
\end{align}
where $\oi^\alpha$ is defined as
\begin{align}
\oi^\alpha&=\int\limits_{t_1}^{t_2} G_i(f(u)+\alpha\eta(u),f^{'}(u)+\alpha{\eta}^{'}(u),u)du,
\end{align}
for $i=1,\ldots,n$. Note that in (\ref{Ch2_Eqn:106}), the order of the derivative with respect to $\alpha$ and the integral can be exchanged due to the Lebesgue dominated convergence theorem \cite{Royden}. Therefore, the Gateaux derivative $\delta_f  L$ can be written as
\begin{align}\nonumber
\delta_f  L=&\int\limits_{t_1}^{t_2} \bigg[\eta(t) F^{f}(f(t),f^{'}(t),\fa ,\ldots, \fn ,t)\\\nonumber
&+{\eta}^{'}(t) F^{f^{'}}(f(t),f^{'}(t),\fa ,\ldots, \fn ,t)\bigg]dt\\\label{Ch2_Eqn:60}
&+\int\limits_{t_1}^{t_2}\sum\limits_{i=1}^{n} \frac{d \oi ^{\alpha}}{d \alpha} \Bigg|_{\alpha=0}\!\!\!\!\!\!\!\!\cdot F^{\oi }(f(t),f^{'}(t),\fa ,\ldots, \fn ,t)dt\\\nonumber
=&\int\limits_{t_1}^{t_2} \eta(t)\bigg[F^{f}(f(t),f^{'}(t),\fa ,\ldots, \fn ,t)\\\nonumber
&-\frac{d}{dt}F^{f^{'}}(f(t),f^{'}(t),\fa ,\ldots, \fn ,t)\bigg]dt\\\nonumber
&+\underbrace{F^{f^{'}}(f(t),f^{'}(t),\fa ,\ldots, \fn ,t)\eta(t)\Bigg|_{t_1}^{t_2}}_{=0}\\\label{Ch2_Eqn:71}
&+\int\limits_{t_1}^{t_2}\sum\limits_{i=1}^{n} \frac{d \oi ^{\alpha}}{d \alpha} \Bigg|_{\alpha=0}\!\!\!\!\!\!\!\!\cdot F^{\oi }(f(t),f^{'}(t),\fa ,\ldots, \fn ,t)dt,
\end{align}
where (\ref{Ch2_Eqn:71}) is due to integration by parts, and the fact that $\eta(t)$ vanishes at $t_1$ and $t_2$ by construction. We compute
\begin{align}
\!\!\!\frac{d\oi ^{\alpha}}{d\alpha}\Bigg|_{\alpha=0}\!\!\!\!\!\!&=\frac{d}{d\alpha}\int\limits_{t_1}^{t_2} G_i(f(u)+\alpha\eta(u),f^{'}(u)+\alpha{\eta}^{'}(u),u)du\Bigg|_{\alpha=0}\!\!\!\!\\\nonumber
&=\int\limits_{t_1}^{t_2} \bigg[\eta(u) G_i^{f}(f(u),f^{'}(u),u)\\
&\quad+{\eta}^{'}(u)G_i^{f^{'}}(f(u),f^{'}(u),u)\bigg]du\\\nonumber
&=\int\limits_{t_1}^{t_2} \eta(u)\bigg[G_i^{f}(f(u),f^{'}(u),u)\\\nonumber
&\quad-\frac{d}{du}G_i^{f^{'}}(f(u),f^{'}(u),u)\bigg]du\\
&\quad+G_i^{f^{'}}(f(u),f^{'}(u),u)\eta(u)\bigg|_{t_1}^{t_2}\\\nonumber
&=\int\limits_{t_1}^{t_2} \eta(u)\bigg[G_i^{f}(f(u),f^{'}(u),u)\\\label{Ch2_Eqn:59}
&\quad-\frac{d}{du}G_i^{f^{'}}(f(u),f^{'}(u),u)\bigg]du,~i=1,\ldots,n.
\end{align}
By plugging (\ref{Ch2_Eqn:59}) into (\ref{Ch2_Eqn:71}) we finally have
\begin{align}\nonumber
\delta_f  L&=\int\limits_{t_1}^{t_2} \eta(u)\bigg[F^{f}(f(u),f^{'}(u),\fa ,\ldots, \fn ,u)\\\nonumber
&\quad\quad-\frac{d}{du}F^{f^{'}}(f(u),f^{'}(u),\fa ,\ldots, \fn ,u)\bigg]du\\\nonumber
&\quad+\sum\limits_{i=1}^{n}\left[\int\limits_{t_1}^{t_2}F^{\oi }(f(t),f^{'}(t),\fa ,\ldots, \fn ,t)dt\right.\\\nonumber
&\quad\quad\quad\quad\cdot\int\limits_{t_1}^{t_2} \eta(u)\bigg(G_i^{f}(f(u),f^{'}(u),u)\\
&\quad\quad\quad\quad\quad-\frac{d}{du}G_i^{f^{'}}(f(u),f^{'}(u),u)\bigg)\left.du \vphantom{\int\limits_{t_1}^{t_2}}\right]\\\nonumber
&=\int\limits_{t_1}^{t_2} \eta(u)\left[\vphantom{\int\limits_{t_1}^{t_2}} F^{f}(f(u),f^{'}(u),\fa ,\ldots, \fn ,u)\right.\\\nonumber
&\quad-\frac{d}{du}F^{f^{'}}(f(u),f^{'}(u),\fa ,\ldots, \fn ,u)\\\nonumber
&\quad+\!\sum\limits_{i=1}^{n}\!\left(\vphantom{\int\limits_{t_1}^{t_2}}\!\!\!\left(\!G_i^{f}(f(u),f^{'}(u),u)\!-\!\frac{d}{du}G_i^{f^{'}}\!(f(u),f^{'}(u),u)\!\right)\right.\\\label{Ch2_Eqn:61}
&\quad\quad\quad\left.\left.\cdot\int\limits_{t_1}^{t_2} F^{\oi }(f(t),f^{'}(t),\fa ,\ldots, \fn ,t)dt\right)\!\right]du.
\end{align}
Since $\eta(u)$ is an arbitrary function and the term multiplying $\eta(u)$ is continuous, it must be zero everywhere on the interval $[t_1,t_2]$. Thus, the optimal solution must satisfy the following equality
\begin{align}\nonumber
&F^{f}(f(u),f^{'}(u),\fa ,\ldots, \fn ,u)\\\nonumber
&-\frac{d}{du}F^{f^{'}}(f(u),f^{'}(u),\fa ,\ldots, \fn ,u)\\\nonumber
&+\!\sum\limits_{i=1}^{n}\!\left(\!\!\vphantom{\int\limits_{t_1}^{t_2}}\left(\!G_i^{f}(f(u),f^{'}(u),u)\!-\!\frac{d}{du}G_i^{f^{'}}(f(u),f^{'}(u),u)\!\right)\right.\\
&\left.\int\limits_{t_1}^{t_2} \!F^{\oi }(f(t),f^{'}(t),\fa ,\ldots, \fn ,t)dt\!\right)\!=\!0,~~\forall u\in[t_1,t_2],
\end{align}
which yields (\ref{Ch2_Eqn:84}).

\QEDA

\begin{rem}\label{Ch2_Remark:4}
When the functional $F$ is independent of $r_i,~i=1,...,n$, then $F^{r_i}(f,f^{'},u)=0$, and therefore, we obtain the well known Euler-Lagrange equation
\begin{align}
F^f(f,f^{'},u)-\frac{d}{du}F^{f^{'}}(f,f^{'},u).
\end{align}
\end{rem}

\begin{rem}\label{Ch2_Remark:1}
As a special case of Lemma (\ref{Ch2_Result:8}), it will be useful in Appendices \ref{Ch2_Appendix:1}, \ref{Ch2_Appendix:2} and \ref{Ch2_Appendix:6} to consider the minimization of the functional
\begin{eqnarray}\label{Ch2_Eqn:26}
L(f)=\frac{1}{\sigma_v}\int\limits_{t_1}^{t_2} \Phi\left(\frac{t}{\sigma_v}\right)\tilde{F}(f(t),\fa,\ldots, \fn ,t)dt,
\end{eqnarray}
where we recall that $\Phi(\cdot)$ is the Gaussian probability density function with mean zero and variance one, and $r_i,~i=1,\ldots,n$, are of the form given by
\begin{equation}
\oi =\frac{1}{\sigma_v}\int\limits_{t_1}^{t_2} \Phi\left(\frac{t}{\sigma_v}\right)\tilde{G}_i(f(t),t)dt,~~i=1,\ldots,n.
\end{equation}
From Lemma \ref{Ch2_Result:8}, setting
\begin{align}
  F&=\frac{1}{\sigma_v}\Phi\left(\frac{t}{\sigma_v}\right)\tilde{F}(f(t),\fa,\ldots, \fn ,t) ,\\
  G_i&=\frac{1}{\sigma_v}\Phi\left(\frac{t}{\sigma_v}\right)\tilde{G}_i(f(t),t),~i=1,...,n,
\end{align}
the solution for this problem needs to satisfy $\nabla L=0$, where the functional derivative $\nabla L$ is found as
\begin{align}\nonumber
\!\!\!\!\nabla L\triangleq&\tilde{F}^{f}(f(u),\fa,\ldots, \fn ,u)+\sum\limits_{i=1}^{n}\left( \vphantom{\int} \tilde{G}_i^{f}(f(u),u)\right.\\\label{Ch2_Eqn:11}
&\left.\!\cdot\!\int\!\frac{1}{\sigma_v}\Phi\!\left(\!\frac{t}{\sigma_v}\!\right)\!\tilde{F}^{\oi }(f(t),\fa,\ldots, \fn ,t)dt\!\right),u\!\in\![t_1,t_2],
\end{align}
with $\tilde{G}_i,~i=1,\ldots,n$, and $\tilde{F}$ being continuous functionals of $(f$, $t)$ and $(f,~\fa,\ldots, \fn,~t)$, respectively, and having continuous partial derivatives with respect to $f$, and $(f$, $\fa,\ldots, \fn)$, respectively.
\end{rem}

\begin{rem}\label{Ch2_Remark:5}
In Appendices \ref{Ch2_Appendix:1}, \ref{Ch2_Appendix:3}, \ref{Ch2_Appendix:5} and \ref{Ch2_Appendix:7} we will consider the minimization of the functional
\begin{align}\label{Ch2_Eqn:27}
L(f)=\frac{1}{\sigma_v}\int\limits_{t_1}^{t_2} \Phi\left(\frac{t}{\sigma_v}\right)\tilde{F}\left(t,f(t)\right)dt.
\end{align}
From Remark \ref{Ch2_Remark:4}, setting
\begin{align}
  F(f,t)&=\frac{1}{\sigma_v}\Phi\left(\frac{t}{\sigma_v}\right)\tilde{F}(f(t),t) ,
\end{align}
the solution for the minimization of the problem (\ref{Ch2_Eqn:27}) needs to satisfy
\begin{align}\label{Ch2_Eqn:28}
\nabla L\triangleq \tilde{F}^{f}\left(u,f(u)\right)=0,\ u\in[t_1,t_2],
\end{align}
where $\tilde{F}$ is continuous with respect to $f$ and has continuous partial derivatives with respect to $f$.
\end{rem}

\section{Proof of Proposition \ref{Ch2_Result:1} } \label{Ch2_Appendix:1}
As discussed in Section \ref{Ch2_Sec:Average Distortion}, the MMSE estimator $g(\cdot)=\mathbb{E}[V|\cdot]$ is optimal for any encoder mapping $f$. Due to the orthogonality principle of the MMSE estimation, we can write $\bar{D}=\sigma_v^2-\mathbb{E}[V\hat{V}]$. Rewriting the Lagrangian in (\ref{Ch2_Eqn:4}) for the MSE distortion criterion, and dropping the constants that are independent of $f$, we have
\begin{align}
L_s(f,g,\lambda)&\triangleq L(f,g,\lambda)-\sigma_v^2\\\label{Ch2_Eqn:36}
&=-\mathbb{E}[V\hat{V}]+\lambda \mathbb{E}[f(V)^2].
\end{align}
In the following, we prove the proposition by means of three key lemmas. In the first, using (\ref{Ch2_Eqn:11}), we obtain a necessary condition for the optimal encoder mapping. Then, using this necessary condition, we show that the Lagrangian function in (\ref{Ch2_Eqn:36}) takes its minimum value when $f$ is odd. Finally, tackling the optimization problem in (\ref{Ch2_Eqn:3}) over odd functions, and using (\ref{Ch2_Eqn:28}), we obtain the result of Proposition \ref{Ch2_Result:1}.

\begin{lem}\label{Ch2_app_lemma:0}
The optimal encoder function $f$ for the problem (\ref{Ch2_Eqn:3}), has to satisfy
\begin{align}\label{Ch2_Eqn:32}
f(v)e^{\frac{f(v)^2}{2\sigma_w^2}}&=a_f(v+b_f),
\end{align}
where $a_f$ and $b_f$ are defined as
\begin{subequations}
\begin{align}
a_f&\triangleq\frac{-\fa }{\sqrt{2\pi}\sigma_w\lambda \fb (1-\fb )},\\
b_f&\triangleq-\frac{(1-2\fb )\fa }{2\fb (1-\fb )},
\end{align}
\end{subequations}
with $r_1$ and $r_2$ defined as
\begin{subequations}
\begin{align}
\fa&\triangleq\frac{1}{\sigma_v}\int v \Phi\left(\frac{v}{\sigma_v}\right)Q\left(\frac{f(v)}{\sigma_w}\right)dv,\\
\fb &\triangleq\frac{1}{\sigma_v}\int \Phi\left(\frac{v}{\sigma_v}\right)Q\left(\frac{f(v)}{\sigma_w}\right)dv.
\end{align}
\end{subequations}
\end{lem}

\textit{Proof}: Expanding the objective function $L_s(f,g,\lambda)$ in (\ref{Ch2_Eqn:36}), we have
\begin{subequations}
\begin{align}\label{Ch2_Eqn:39}
&L_s(f,g,\lambda)=-\mathbb{E}_V\left[V\mathbb{E}_{\hat{V}|V}[\hat{V}|V]-\lambda f(V)^2\right]\\
&=\!\frac{1}{\sigma_v}\!\int\! \Phi\!\left(\!\frac{v}{\sigma_v}\!\right)\!\left(\!vQ\!\left(\!\frac{f(v)}{\sigma_w}\!\right)\!\left(\hat{v}_{(1)}-\hat{v}_{(2)}\right)\!+\!\lambda f(v)^2\!\right)\! dv\!\\\label{Ch2_Eqn:29}
&=\!\frac{1}{\sigma_v}\!\int \! \Phi\!\left(\!\frac{v}{\sigma_v}\!\right)\!\left(\!vQ\!\left(\!\frac{f(v)}{\sigma_w}\!\right)\!\left(\!\frac{\fa}{\fb (\fb-1)}\!\right)\!+\!\lambda f(v)^2\!\right)\!dv,
\end{align}
\end{subequations}
where $\hat{v}_{(1)}=\frac{\fa}{\fb-1}$ and $\hat{v}_{(2)}=\frac{\fa}{\fb }$. We observe that (\ref{Ch2_Eqn:29}) can be stated in the form in (\ref{Ch2_Eqn:26}) by setting
\begin{subequations}
\begin{align}
  \tilde{F} &= vQ\left(\frac{f(v)}{\sigma_w}\right)\cdot\left(\frac{\fa}{\fb (\fb-1)}\right)+\lambda f(v)^2, \\
  \tilde{G}_1&=vQ\left(\frac{f(v)}{\sigma_w}\right),\\
  \tilde{G}_2&=Q\left(\frac{f(v)}{\sigma_w}\right).
\end{align}
\end{subequations}
Note that $\nabla L=\nabla L_s$. Therefore, from (\ref{Ch2_Eqn:11}) we have the necessary condition
\begin{align}\nonumber
\!\!\!\!\!\nabla L&=\frac{v\fa e^{-\frac{f(v)^2}{2\sigma_w^2}}}{\sqrt{2\pi}\sigma_w\fb(1-\fb)}+2\lambda f(v)\\\nonumber
&\quad+\frac{v \fa e^{-\frac{f(v)^2}{2\sigma_w^2}}}{\sqrt{2\pi}\sigma_w\fb(1-\fb)}-\frac{(1-2\fb)\fa^2 }{(\fb(1-\fb))^2} \frac{e^{-\frac{f(v)^2}{2\sigma_w^2}}}{\sqrt{2\pi}\sigma_w}\\\label{Ch2_Eqn:31}
&=\!\frac{2\fa e^{-\frac{f(v)^2}{2\sigma_w^2}}}{\sqrt{2\pi}\sigma_w\fb (1-\fb )}\!\left(v\!-\!\frac{(1-2\fb )\fa }{2\fb (1-\fb )}\right)\!+\!2\lambda f(v)\!=\!0.
\end{align}
By solving (\ref{Ch2_Eqn:31}) with respect to $f(v)$, we obtain (\ref{Ch2_Eqn:32}), which concludes the proof.

\qed

Based on (\ref{Ch2_Eqn:32}), we restrict the minimization of the objective function over encoder mappings $f_{a,b}(v)$ that satisfy
\begin{align}\label{Ch2_Eqn:35}
f_{a,b}(v)e^{\frac{f_{a,b}(v)^2}{2\sigma_w^2}}=\frac{1}{a}(v-b),
\end{align}
for some parameters $a$ and $b$. Note that taking the derivative of (\ref{Ch2_Eqn:35}) with respect to $b$ we have
\begin{align}\label{Ch2_Eqn:104}
f_{a,b}^{'}(v)=\frac{1}{ae^{\frac{f_{a,b}(v)^2}{2\sigma_w^2}}\left(1+2f_{a,b}(v)^2\right)},
\end{align}
which implies that for $a>0$, the function $f_{a,b}(v)$ is monotonically increasing.

\begin{figure}
\begin{centering}
\includegraphics[scale=0.47]{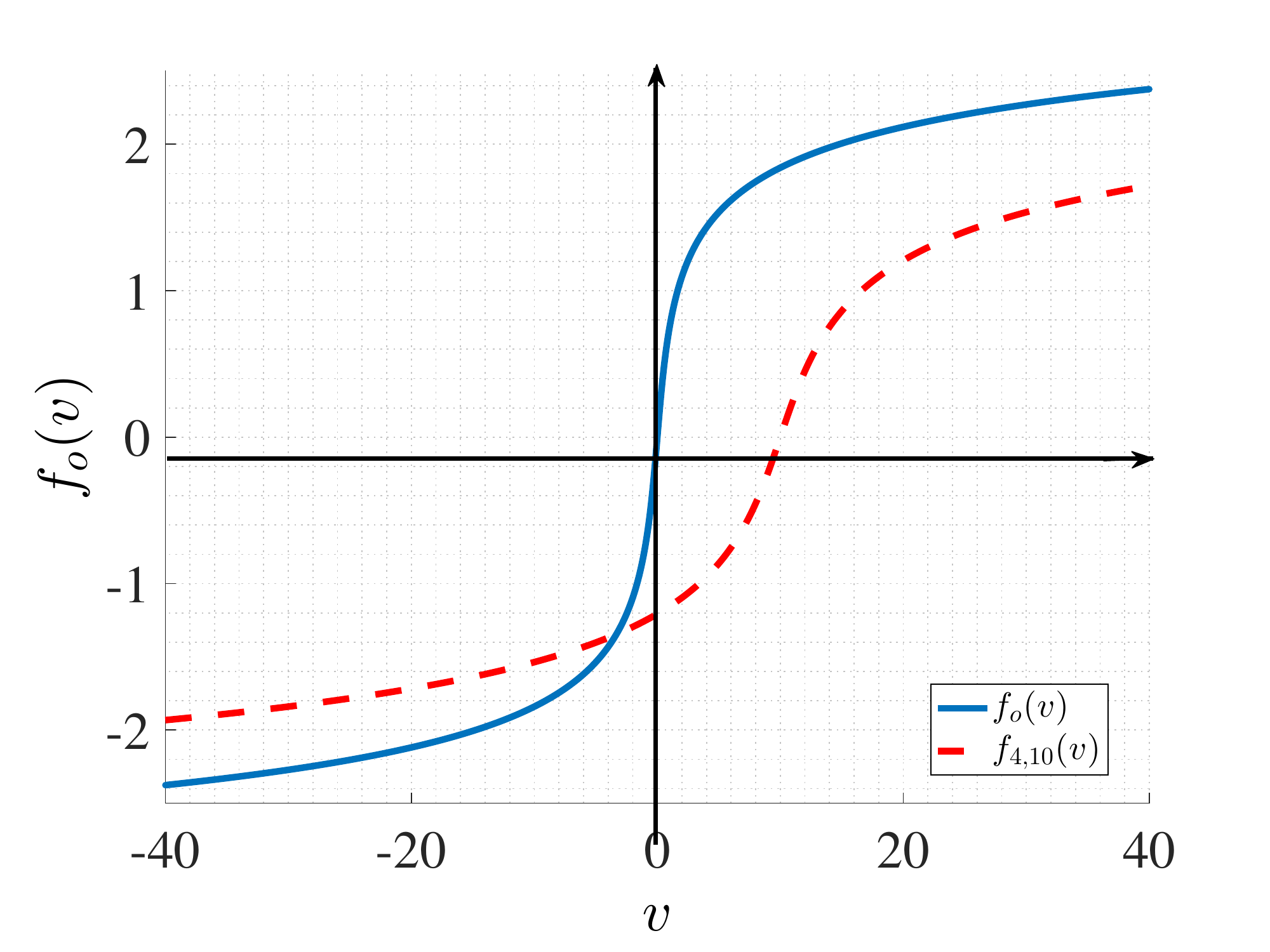}
\caption{The function $f_o(v)$ in (\ref{Ch2_Eqn:33}) and the function $f_{a,b}(v)$ in (\ref{Ch2_Eqn:35}) with $a=4,~b=10$ .}\label{Ch2_Fig:5}
\par\end{centering}
\vspace{0mm}
\end{figure}

Defining the function $f_o(\cdot)$ as the unique solution of
\begin{align}\label{Ch2_Eqn:33}
f_o(v)e^{\frac{f_o(v)^2}{2}}=v,
\end{align}
it will be convenient to write
\begin{align}
f_{a,b}(v)=\sigma_wf_o\left(\frac{v-b}{a\sigma_w}\right).
\end{align}
This shows that any function $f_{a,b}(v)$ can be seen as a scaled and shifted version of the function that satisfies (\ref{Ch2_Eqn:33}). The function $f_o(v)$ is plotted in Figure \ref{Ch2_Fig:5}, along with an example of the function $f_{a,b}(v)$ for $a=4,~b=10$. It can be easily verified that $f_o(v)$ is odd, that is, $f_o(-v)=-f_o(v)$. Furthermore, functions of the form $f_{a,0}(v)$, that is, with $b=0$, are odd as well.

In the following lemma, we show that the Lagrangian functional in (\ref{Ch2_Eqn:36}) takes its minimum value only over functions of the form $f_{a,b}(v)$ where $b=0$ for any $a$, which shows that the optimal function $f_{a,b}(v)$ is odd as summarized in Corollary \ref{Ch2_Cor:1}.

\begin{lem}\label{Ch2_app_lemma:1}
The optimal solution of the problem
\begin{eqnarray}\label{ch2_Eqn:108}
\begin{aligned}
& \underset{b}{\text{minimize}}
& & L(f_{a,b},g,\lambda),
\end{aligned}
\end{eqnarray}
is achieved for $b=0$ for any fixed $a$.
\end{lem}
\textit{Proof}: We prove the lemma assuming that $a,b\geq 0$. This is without loss of generality given the relationship $L(f_{a,b},g,\lambda)=L(f_{a,-b},g,\lambda)=L(f_{-a,b},g,\lambda)=L(f_{-a,-b},g,\lambda)$. This can be verified by substituting $f_{a,-b},~f_{-a,b}$ and $f_{-a,-b}$ in (\ref{Ch2_Eqn:36}) and performing the changes of variables $v=t-2b,~v=2b-t$ and $v=-t$, respectively. In the following, we show that decreasing the value of $b\geq 0$, reduces the average power of the mapping, i.e., $\mathbb{E}[f_{a,b}(V)^2]$, and increases $\mathbb{E}[V\hat{V}]$. Therefore, the function $L(f_{a,b},g,\lambda)$ becomes smaller by decreasing the value of $b\geq 0$.

\textit{1) $\mathbb{E}[f_{a,b}(V)^2]$ is a strictly increasing function of $b$:} By writing the average power of the function $f_{a,b}(v)$ we have
\begin{align}
\mathbb{E}[f_{a,b}(V)^2]&=\frac{1}{\sigma_v}\int\Phi\left(\frac{v}{\sigma_v}\right)f_{a,b}(v)^2dv\\
&=\frac{\sigma_w}{\sigma_v}\int\Phi\left(\frac{v}{\sigma_v}\right)f_o\left(\frac{v-b}{a\sigma_w}\right)^2dv\\
&=\frac{a\sigma_w^2}{\sigma_v}\int\Phi\left(\frac{a\sigma_w v+b}{\sigma_v}\right)f_o\left(v\right)^2dv.
\end{align}
Differentiating $\mathbb{E}[f_{a,b}(V)^2]$ with respect to $b$, we have
\begin{align}\nonumber
-&\frac{d\mathbb{E}[f_{a,b}(V)^2]}{db}\\
&=\frac{a\sigma_w^2}{\sigma_v^3}\int(a\sigma_wv+b)\Phi\left(\frac{a\sigma_w v+b}{\sigma_v}\right)f_o\left(v\right)^2dv\\\label{Ch2_Eqn:38}
&\leq \frac{a\sigma_w^2}{\sigma_v^3} \int(a\sigma_wv+b)\Phi\left(\frac{a\sigma_w v+b}{\sigma_v}\right)\Psi(v)dv,
\end{align}
where $\Psi(v)$ is defined as
\begin{align}
\Psi(v)\triangleq\left\{\begin{array}{cc}
         f_o(v)^2 & v\leq -\frac{b}{a\sigma_w}, \\
         f_o\left(v+\frac{2b}{a\sigma_w}\right)^2 & v> -\frac{b}{a\sigma_w}.
       \end{array}\right.
\end{align}

It is noted that the inequality in (\ref{Ch2_Eqn:38}) holds due to the fact that the integrand is positive and that we have the inequalities $\Psi(v)\geq f_o(v)$ for $v\geq -b/(a\sigma_w) $. Also, $\Psi(v)$ is an even function shifted to the left by $\frac{b}{a\sigma_w}$. Hence, due to the fact that $(a\sigma_w v + b)\phi(\frac{a\sigma_w v + b}{\sigma_v})$ is odd with respect to $v=-b/(a\sigma_w )$, the integrand on the right hand side of the inequality in (\ref{Ch2_Eqn:38}) is an odd function shifted to the left by $b/(a\sigma_w$.  Therefore, the integral is zero, completing the proof.  Note that (\ref{Ch2_Eqn:38}) holds with equality only when $b=0$.

\textit{2) $\mathbb{E}[V\hat{V}]$ is a decreasing function of $b$:}
We first consider the noiseless scenario, i.e., $W=0$. In the noiseless scenario, for the reconstruction points we have
\begin{align}
\hat{V}&=\left\{\begin{array}{ll}
\mathbb{E}[V|v\geq b]&Y= 0,\\
\mathbb{E}[V|v< b]&Y= 1
\end{array}\right.\\
&=\left\{\begin{array}{ll}
\frac{\sigma_v e^{-\frac{b^2}{2}}}{\sqrt{2\pi}Q\left(\frac{b}{\sigma_v}\right)}&Y= 0, \vspace{2mm}\\
\frac{-\sigma_v e^{-\frac{b^2}{2}}}{\sqrt{2\pi}\left(1-Q\left(\frac{b}{\sigma_v}\right)\right)}&Y= 1.
\end{array}\right.
\end{align}
Therefore, $\mathbb{E}[V\hat{V}]$ can be written as
\begin{align}\nonumber
\mathbb{E}[V\hat{V}]&=\textrm{Pr}(V\geq b)\hat{v}_{(1)}\mathbb{E}[V|\hat{V}=\hat{v}_{(1)}]\\
&\quad+\textrm{Pr}(V< b)\hat{v}_{(2)}\mathbb{E}[V|\hat{V}=\hat{v}_{(2)}]\\\label{Ch2_Eqn:41}
&=\frac{\sigma_v^2 e^{-b^2}}{2\pi Q\left(\frac{b}{\sigma_v}\right)\left(1-Q\left(\frac{b}{\sigma_v}\right)\right)}\triangleq \tilde{g}(b).
\end{align}
It can be verified that the function $\tilde{g}(b)$ is even, and that it takes its maximum at $b=0$ as seen in Figure \ref{Ch2_Fig:8} (for the proof see Appendix \ref{Ch2_Appendix:10}).

Now, considering the noisy received signal, we expand $\mathbb{E}[V\hat{V}]$ as
\begin{align}
\mathbb{E}[V\hat{V}]&=\frac{1}{\sigma_w}\int \Phi\left(\frac{w}{\sigma_w}\right) \mathbb{E}[V\hat{V}|W=-w] dw,
\end{align}
where
\begin{align}\nonumber
\mathbb{E}&[V\hat{V}|W=-w]\\\nonumber
&=\mathbb{E}[V|Y=0,W=-w]^2\mathrm{Pr}(Y=0|W=-w)\\
&\quad+\mathbb{E}[V|Y=1,W=-w]^2\mathrm{Pr}(Y=1|W=-w)\\
&=\frac{\sigma_ve^{-\left(f_{a,b}^{-1}(w)\right)^2}}{2\pi Q\left(\frac{f_{a,b}^{-1}(w)}{\sigma_v}\right)}+\frac{\sigma_ve^{-\left(f_{a,b}^{-1}(w)\right)^2}}{2\pi \left(1-Q\left(\frac{f_{a,b}^{-1}(w)}{\sigma_v}\right)\right)}\\
&=\tilde{g}\left(f_{a,b}^{-1}(w)\right).
\end{align}
Notice that the inverse of the function $f_{a,b}(v)$ exists because it is one-to-one and is defined on the whole real line. Therefore,
\begin{align}
\mathbb{E}[V\hat{V}]&=\frac{1}{\sigma_w}\int \Phi\left(\frac{w}{\sigma_w}\right) \tilde{g}\left(f_{a,b}^{-1}(w)\right) dw.
\end{align}

We want to show that
\begin{align}\label{Ch2_Eqn:43}
\int e^{-\frac{w^2}{2}} \tilde{g}\left(f_{a,b}^{-1}(w)\right) dw\leq \int e^{-\frac{w^2}{2}} \tilde{g}\left(f_{a,0}^{-1}(w)\right) dw.
\end{align}
Using the change of variables $w=f_{a,0}(v)$ and the equality $f_{a,b}^{-1}(w)=b+f_{a,0}^{-1}(w)$, inequality (\ref{Ch2_Eqn:43}) can be written as
\begin{align}\label{Ch2_Eqn:34}
\int e^{-\frac{f_{a,0}^2(v)}{2}} \tilde{g}(b+v) f^{'}_{a,0}(v)dv\leq \int e^{-\frac{f_{a,0}^2(v)}{2}} \tilde{g}(v) f^{'}_{a,0}(v)dv.
\end{align}
Inequality (\ref{Ch2_Eqn:34}) is equivalent to
\begin{align}\nonumber
\int\limits_{-\infty}^{-\frac{b}{2}}& e^{-\frac{f_{a,0}^2(v)}{2}} \left(\tilde{g}(b+v)-\tilde{g}(v)\right) f^{'}_{a,0}(v)dv\leq\\\label{Ch2_Eqn:44}
 &\int\limits_{-\frac{b}{2}}^{\infty} e^{-\frac{f_{a,0}^2(v)}{2}} \left(\tilde{g}(v)-\tilde{g}(b+v)\right) f^{'}_{a,0}(v)dv.
\end{align}
Using the transformation $-b-v=t$ for the left hand side of (\ref{Ch2_Eqn:44}), we have
\begin{align}\nonumber
\int\limits_{-\frac{b}{2}}^{\infty}&e^{-\frac{f_{a,0}^2(-b-t)}{2}} \left(\tilde{g}(-t)-\tilde{g}(-b-t)\right) f^{'}_{a,0}(-t-b)dt\\
&=\int\limits_{-\frac{b}{2}}^{\infty} e^{-\frac{f_{a,0}^2(b+t)}{2}} \left(\tilde{g}(t)-\tilde{g}(b+t)\right) f^{'}_{a,0}(t+b)dt,
\end{align}
where the second equality is due to the fact that functions $\tilde{g}$, $f^2_{a,0}$, $f^{'}_{a,0}$ are even. Therefore, (\ref{Ch2_Eqn:44}) is equivalent to
\begin{align}\nonumber
\int\limits_{-\frac{b}{2}}^{\infty} &\left[e^{-\frac{f_{a,0}^2(b+t)}{2}} f^{'}_{a,0}(t+b)-e^{-\frac{f_{a,0}^2(t)}{2}} f^{'}_{a,0}(t)\right]\\\label{Ch2_Eqn:45}
&\quad\quad\quad\cdot\left[\tilde{g}(t)-\tilde{g}(b+t)\right]dt\leq 0.
\end{align}
The inequality in (\ref{Ch2_Eqn:45}) concludes the proof and follows from the following facts:

\begin{figure}
\begin{centering}
\includegraphics[scale=0.47]{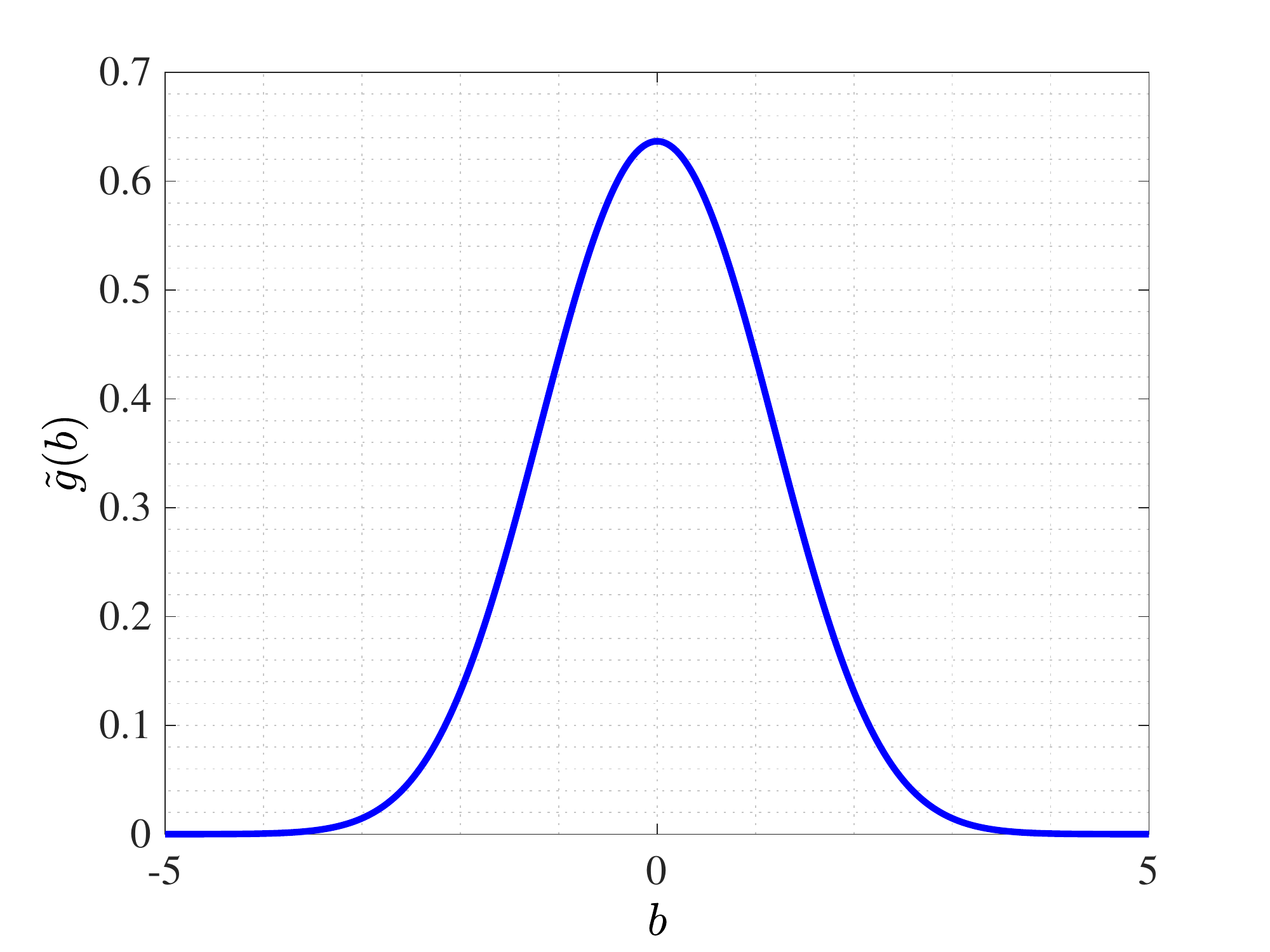}
\caption{Plot of the function $\tilde{g}(b)$ in (\ref{Ch2_Eqn:41}).}\label{Ch2_Fig:8}
\par\end{centering}
\vspace{0mm}
\end{figure}

\begin{itemize}
        \item Since $g(t)$ is even and decreasing for $t>0$, we have $\tilde{g}(t)\geq \tilde{g}(t+b)>0$ for $t\geq -\frac{b}{2}$, and hence the second bracket is nonnegative;
        \item Since $f_{a,0}(\cdot)$ is nondecreasing, we have the inequality $e^{-\frac{f_{a,0}^2(t)}{2}} \geq e^{-\frac{f_{a,0}^2(b+t)}{2}}$;
        \item From (\ref{Ch2_Eqn:104}) given that the function $f^{'}_{a,0}(t)$ is even and decreasing for $t>-\frac{b}{2}$, we have the inequality $f^{'}_{a,0}(t)\geq f^{'}_{a,0}(t+b)>0$ for $t>-\frac{b}{2}$.
      \end{itemize}
      Therefore, the inequality in (\ref{Ch2_Eqn:43}) is concluded.

\qed

\begin{cor}\label{Ch2_Cor:1}
  The optimal mapping for problem (\ref{Ch2_Eqn:4}) under the MSE criterion is odd.

  \textit{Proof}: By Lemma \ref{Ch2_app_lemma:0}, the optimal mapping can be written in the form $f_{a,b}(v)$ without loss of optimality. By Lemma \ref{Ch2_app_lemma:1}, we conclude that the optimal mapping is in the form $f_{a,0}(v)$; and hence, it is odd.
\end{cor}

\begin{lem}
A necessary condition for an optimal solution to problem (\ref{Ch2_Eqn:3}) is
\begin{align}\label{Ch2_Eqn:48}
\nabla L=\frac{-2v}{\sqrt{2\pi}\sigma_w}e^{-\frac{f(v)^2}{2\sigma_w^2}}+2\lambda f(v)=0.
\end{align}
\end{lem}

\textit{Proof}: Since the optimal mapping is odd by Corollary \ref{Ch2_Cor:1}, it can be easily verified that $\hat{v}_{(1)}=-\hat{v}_{(2)}$. Furthermore, without loss of optimality we can assume $\hat{v}_{(1)}\geq 0$. Therefore, using (\ref{Ch2_Eqn:8}) problem (\ref{Ch2_Eqn:3}) can be restated as
\begin{equation}\label{Ch2_Eqn:46}
\begin{aligned}
& \underset{f}{\text{minimize}}
& & -\hat{v}_{(1)}+\lambda E[f(V)^2],
\end{aligned}
\end{equation}
where $\hat{v}_{(1)}$ is obtained as in (\ref{Ch2_Eqn:7}). Expanding (\ref{Ch2_Eqn:46}), we can write
\begin{align}\label{Ch2_Eqn:47}
& \underset{f}{\text{minimize}}\quad
\!\!\frac{-1}{\sigma_v}\!\int\limits_{-\infty}^{\infty}\!\Phi\!\left(\!\frac{v}{\sigma_v}\!\right)\!\left(2vQ\!\left(\!\frac{f(v)}{\sigma_w}\!\right)\!-\!\lambda f(v)^2\!\right)\!dv.
\end{align}
Since (\ref{Ch2_Eqn:47}) is of the form (\ref{Ch2_Eqn:27}) with
\begin{align}
\tilde{F}=-2vQ\left(\frac{f(v)}{\sigma_w}\right)+\lambda f(v)^2,
\end{align}
we have the necessary condition in (\ref{Ch2_Eqn:48}) using (\ref{Ch2_Eqn:28}). This concludes the proof.

\QEDA

\section{Proof of Proposition \ref{Ch2_Result:2} } \label{Ch2_Appendix:2}
Expanding the objective function $L(f,g,\lambda)$ as in (\ref{Ch2_Eqn:39}) we have
\begin{subequations}
\begin{align}\nonumber
&L_s(f,g,\lambda)\\
&=\!\frac{-1}{\sigma_w}\!\int\!\! \Phi\!\left(\!\frac{v}{\sigma_v}\!\right)\!\!\left(\! v\! \!\int \! \! g\left(Q\left(f(v)+w\right)\right)\Phi\!\left(\!\frac{w}{\sigma_w}\!\right)\!dw\!-\!\lambda f(v)^2\!\!\vphantom{\int} \right) \!dv\\\nonumber
&=\!\frac{-1}{\sigma_v}\!\int\! \Phi\!\left(\!\frac{v}{\sigma_v}\!\right) \! \left(\!v\sum\limits_{j=1}^{K}\hat{v}_{(j)}\!\left(\!Q\!\left(\!\frac{z_{(j-1)}\!-\!f(v)}{\sigma_w}\!\right)\!\right.\right.\\\label{Ch2_Eqn:76}
&\left.\left.\quad\quad\quad\quad\quad\quad- Q\left(\!\frac{z_{(j)}\!-\!f(v)}{\sigma_w}\!\right)\!\right)\!-\!\lambda f(v)^2\!\vphantom{\sum\limits_{j=1}^{K}} \right)\!dv,
\end{align}
\end{subequations}
where we recall that $\hat{v}_{(j)}$ is the MMSE estimation of the source when the received signal is $Y=y_{(j)}$, i.e., $Y=f(V)+W\in[z_{(j-1)},z_{(j)})$. We have $\hat{v}_{(j)}=\frac{r_{1j}}{r_{2j}}$ with
\begin{subequations}
\begin{align}
r_{1j}\! &=\!\frac{1}{\sigma_v}\!\int\! \! v\Phi\!\left(\!\frac{v}{\sigma_v}\!\right)\!\!\left(\!Q\!\left(\!\frac{z_{(j-1)}\!-\!f(v)}{\sigma_w}\!\right)\!\!-\!Q\!\left(\!\frac{z_{(j)}\!-\!f(v)}{\sigma_w}\!\right)\!\right)\!dv,\\
r_{2j} \!&=\!\frac{1}{\sigma_v}\!\int \!\! \Phi\!\left(\!\frac{v}{\sigma_v}\!\right)\!\!\left(\!Q\!\left(\!\frac{z_{(j-1)}\!-\!f(v)}{\sigma_w}\!\right)\!\!-\!Q\!\left(\!\frac{z_{(j)}\!-\!f(v)}{\sigma_w}\!\right)\!\right)\!dv,
\end{align}
\end{subequations}
for $j=1,...,K$. Since (\ref{Ch2_Eqn:76}) is of the form (\ref{Ch2_Eqn:26}) with
\begin{align}\nonumber
\tilde{F}\!&=\!-v\sum\limits_{j=1}^{K}\frac{r_{1j}}{r_{2j}}\!\left(\!Q\!\left(\!\frac{z_{(j-1)}\!-\!f(v)}{\sigma_w}\!\right)\!-\!Q\!\left(\!\frac{z_{(j)}\!-\!f(v)}{\sigma_w}\!\right)\!\right)\!\\
&\quad+\lambda f(v)^2,
\end{align}
and
\begin{align}\nonumber
\tilde{G}_{1j}\!&=\!v\!\left(\!Q\!\left(\!\frac{z_{(j-1)}\!-\!f(v)}{\sigma_w}\!\right)\!-\!Q\!\left(\!\frac{z_{(j)}\!-\!f(v)}{\sigma_w}\!\right)\!\right),\\
\tilde{G}_{2j}\!&=\!\left(\!Q\!\left(\!\frac{z_{(j-1)}\!-\!f(v)}{\sigma_w}\!\right)\!-\!Q\!\left(\!\frac{z_{(j)}\!-\!f(v)}{\sigma_w}\!\right)\!\right),
\end{align}
for $j=1,...,K$. By writing the necessary condition in (\ref{Ch2_Eqn:11}), we have
\begin{align}\nonumber
\nabla L&=-\frac{v}{\sqrt{2\pi}\sigma_w}\sum\limits_{j=1}^{K}\frac{r_{1j}}{r_{2j}}\!\left(\!e^{-\frac{\left(z_{(j-1)}\!-\!f(v)\right)^2}{2\sigma_w^2}}\!-\!e^{-\frac{\left(z_{(j)}\!-\!f(v)\right)^2}{2\sigma_w^2}}\!\right)\\\nonumber
&\quad-\frac{v}{\sqrt{2\pi}\sigma_w}\sum\limits_{j=1}^{K}\frac{r_{1j}}{r_{2j}}\!\left(\!e^{-\frac{\left(z_{(j-1)}\!-\!f(v)\right)^2}{2\sigma_w^2}}\!-\!e^{-\frac{\left(z_{(j)}\!-\!f(v)\right)^2}{2\sigma_w^2}}\!\right)\\\nonumber
&\quad+\frac{1}{\sqrt{2\pi}\sigma_w}\sum\limits_{j=1}^{K}\frac{r_{1j}^2}{r_{2j}^2}\!\left(\!e^{-\frac{\left(z_{(j-1)}\!-\!f(v)\right)^2}{2\sigma_w^2}}\!-\!e^{-\frac{\left(z_{(j)}\!-\!f(v)\right)^2}{2\sigma_w^2}}\!\right)\\\label{Ch2_Eqn:49}
&\quad+2\lambda f(v)=0.
\end{align}
Solving for $f$, we have
\begin{align} \nonumber
f(v)&=\frac{1}{2\sqrt{2\pi}\sigma_w\lambda}\sum\limits_{j=1}^{K}\left(\vphantom{e^{-\frac{\left(z_{(j-1)}-f(v)\right)^2}{2\sigma_w^2}}} \hat{v}_{(j)}\left(2v-\hat{v}_{(j)}\right)\right.\\
&\left.\quad\quad\quad\cdot\left(e^{-\frac{\left(z_{(j-1)}-f(v)\right)^2}{2\sigma_w^2}}-e^{-\frac{\left(z_{(j)}-f(v)\right)^2}{2\sigma_w^2}}\right)\right).
\end{align}

\QEDA

\section{Proof of Proposition \ref{Ch2_Result:3} } \label{Ch2_Appendix:3}
We start by expanding the Lagrangian function (\ref{Ch2_Eqn:4}) for the DOP criterion as
\begin{subequations}\label{Ch2_Eqn:85}
\begin{align}
L(f,g,\lambda)&=\epsilon(D)+\lambda \mathbb{E}[f(V)^2]\\\nonumber
&=\mathrm{Pr}\left(V\in (I_1\cup I_2)^{C}\right)\\\nonumber
&+\mathrm{Pr}\left(V\in (I_1\setminus I_2),\hat{V}=\hat{v}_{(2)}\right)\\\nonumber
&+\mathrm{Pr}\left(V\in (I_2\setminus I_1),\hat{V}=\hat{v}_{(1)}\right)\\\label{Ch2_Eqn:79}
&\quad+\frac{\lambda}{\sigma_v}\int\Phi\left(\frac{v}{\sigma_v}\right)f(v)^2dv,
\end{align}
\end{subequations}
where we have used the decomposition in (\ref{Ch2_Eqn:17}). The probabilities in (\ref{Ch2_Eqn:79}) can be written as
\begin{subequations}\label{Ch2_Eqn:50}
\begin{align}
&\mathrm{Pr}\left(V\in (I_1\cup I_2)^{C}\right)=\frac{1}{\sigma_v}\int\limits_{v\in(I_1\cup I_2)^{C}}\Phi\left(\frac{v}{\sigma_v}\right)dv,\\\nonumber
&\mathrm{Pr}\left(V\in (I_1\setminus I_2),\hat{V}=\hat{v}_{(2)}\right)\\
&\quad\quad\quad\quad\quad=\frac{1}{\sigma_v}\int\limits_{v\in I_{1}\setminus I_2}\Phi\left(\frac{v}{\sigma_v}\right)Q\left(\frac{f(v)}{\sigma_w}\right)dv,\\\nonumber
&\mathrm{Pr}\left(V\in (I_2\setminus I_1),\hat{V}=\hat{v}_{(1)}\right)\\
&\quad\quad\quad\quad\quad=\frac{1}{\sigma_v}\int\limits_{v\in I_{2}\setminus I_1}\Phi\left(\frac{v}{\sigma_v}\right)Q\left(\frac{-f(v)}{\sigma_w}\right)dv.
\end{align}
\end{subequations}
Since the intervals on which the integrals in (\ref{Ch2_Eqn:50}) are taken do not overlap and span the real line, the Lagrangian in (\ref{Ch2_Eqn:85}) can be written in the form of (\ref{Ch2_Eqn:27}) with $\tilde{F}(v,f)$ defined as
\begin{align}\label{Ch2_Eqn:110}
\tilde{F}\left(v,f\right)&=\left\{\begin{array}{ll}
1+\lambda f(v)^2& v\in (I_1\cup I_2)^{C},\\
Q\left(\frac{f(v)}{\sigma_w}\right)+\lambda f(v)^2&  v\in (I_1\setminus I_2),\\
Q\left(\frac{-f(v)}{\sigma_w}\right)+\lambda f(v)^2&  v\in (I_2\setminus I_1),\\
\lambda f(v)^2& v\in (I_1\cap I_2).
\end{array}\right.
\end{align}
Using the optimality condition in (\ref{Ch2_Eqn:28}) and differentiating $\tilde{F}(v,f)$ in (\ref{Ch2_Eqn:110}) with respect to $f$, we have the necessary condition $\tilde{F}^f=0$ with
\begin{align}\nonumber
&\tilde{F}^f\left(v,f\right)=\\\label{Ch2_Eqn:51}
&\left\{\begin{array}{ll}
2\lambda f(v)& v\in (I_1\cup I_2)^{C}\cup (I_1\cap I_2), \\
\frac{-1}{\sqrt{2\pi}\sigma_w}e^{\frac{-f(v)^2}{2\sigma_w^2}}+2\lambda f(v)&  v\in (I_1\setminus I_2),\\
\frac{1}{\sqrt{2\pi}\sigma_w}e^{\frac{-f(v)^2}{2\sigma_w^2}}+2\lambda f(v)&   v\in (I_2\setminus I_1).\\
\end{array}\right.
\end{align}
Therefore, we have
\begin{align}
f(v)&=\left\{\begin{array}{ll}
0& v\in (I_1\cup I_2)^{C}\cup (I_1\cap I_2), \\
\frac{1}{2\lambda\sqrt{2\pi}\sigma_w}e^{\frac{-f(v)^2}{2\sigma_w^2}}&  v\in (I_1\setminus I_2),\\
\frac{-1}{2\lambda\sqrt{2\pi}\sigma_w}e^{\frac{-f(v)^2}{2\sigma_w^2}}&   v\in (I_2\setminus I_1).\\
\end{array}\right.
\end{align}

\QEDA

\section{Proof of Proposition \ref{Ch2_Result:4} } \label{Ch2_Appendix:4}
Given $\lambda$, and obtaining the value of $u$ from (\ref{Ch2_Eqn:19}), we aim to minimize the objective function in (\ref{Ch2_Eqn:4}) for the DOP with respect to the reconstruction points $\hat{v}_{(1)}$ and $\hat{v}_{(2)}$. Since by Proposition \ref{Ch2_Result:3} the non-zero optimal values of $f(v)$ are opposite of one another over $I_{2}\setminus I_1$ and $I_{1}\setminus I_2$, we can rewrite (\ref{Ch2_Eqn:4}) for the DOP as follows
\begin{align}\nonumber
&L(f,g,\lambda)=\frac{1}{\sigma_v}\left(\int\limits_{(I_1\cup I_2)^{C}}\Phi\left(\frac{v}{\sigma_v}\right)dv\right.\\\nonumber
&\left.+Q\left(\frac{u}{\sigma_w}\right)\left(\int\limits_{  I_{2}\setminus I_1}\Phi\left(\frac{v}{\sigma_v}\right)
dv+\int\limits_{  I_{1}\setminus I_2}\Phi\left(\frac{v}{\sigma_v}\right)dv \right)\right.\\
&\left.+\lambda \left(\int\limits_{  I_{2}\setminus I_1}\Phi\left(\frac{v}{\sigma_v}\right) u^2 dv+\int\limits_{  I_{1}\setminus I_2}\Phi\left(\frac{v}{\sigma_v}\right) u^2 dv\right)\right)\\\nonumber
&=\frac{1}{\sigma_v}\left(\int\limits_{(I_1\cup I_2)^{C}}\Phi\left(\frac{v}{\sigma_v}\right)dv\right.\\\label{Ch2_Eqn:52}
&\left.+\left(Q\left(\frac{u}{\sigma_w}\right)+\lambda u^2\right)\cdot\int\limits_{  (I_{2}\setminus I_1) \cup  (I_{1}\setminus I_2)}\Phi\left(\frac{v}{\sigma_v}\right)dv\right),
\end{align}
where $u$ is the solution of $ue^{\frac{u^2}{2\sigma_w^2}}=\frac{1}{2\sigma_w\sqrt{2\pi}\lambda}$. From (\ref{Ch2_Eqn:52}), it can be shown that (\ref{Ch2_Eqn:4}) for the DOP is minimized if $I_{1}$ and $I_{2}$ lie around the origin, i.e., $0\in I_1\cap I_2$, and they are symmetric with respect to the origin; that is, if $I_1=[l,h]$ for $l\leq 0 \leq h$, then $I_2=[-h,-l]$. From (\ref{Ch2_Eqn:19}), we have $\lambda u^2=\frac{ue^{\frac{-u^2}{2\sigma_w^2}}}{2\sigma_w\sqrt{2\pi}}$. By substituting this into (\ref{Ch2_Eqn:52}) we have
      \begin{align}\nonumber
      L(f,g,\lambda)&=\frac{1}{\sigma_v}\left(\int\limits_{(I_1\cup I_2)^{C}}\Phi\left(\frac{v}{\sigma_v}\right)dv\right.\\
      &\left.\quad\quad\quad+C(u)\cdot\!\!\!\!\!\!\!\!\!\!\int\limits_{  (I_{2}\setminus I_1) \cup  (I_{1}\setminus I_2)}\!\!\!\!\!\!\Phi\left(\frac{v}{\sigma_v}\right)dv\right),
      \end{align}
      where $C(u)\triangleq Q\left(\frac{u}{\sigma_w}\right)+\frac{ue^{\frac{-u^2}{2\sigma_w^2}}}{2\sigma_w\sqrt{2\pi}}$. By differentiating $C(u)$ with respect to $u$ it can be verified that $C(u)$ is decreasing with $u$ and is less than or equal to $1/2$ for all $u\geq 0$. We now prove the claim (the intervals are around the origin and symmetric) for the case in which the intervals $I_1$ and $I_2$ do not overlap. The proof is similar for the case in which the intervals overlap.

      Consider two symmetric non-overlapping intervals $I_1^s$ and $I_2^s$, such that $I_1^s=[-2\sqrt{D},0]$ and $I_2^s=[0,2\sqrt{D}]$. Note that we  have
      \begin{align}\nonumber
      \frac{1}{\sigma_v}\int\limits_{  I_{1} \cup  I_{2}}&\Phi\left(\frac{v}{\sigma_v}\right)dv\\
      &\leq \frac{1}{\sigma_v}\int\limits_{I_{1}^s \cup  I_{2}^s}\Phi\left(\frac{v}{\sigma_v}\right)dv\\
      &=\frac{1}{\sigma_v}\int\limits_{  I_{1} \cup  I_{2}}\Phi\left(\frac{v}{\sigma_v}\right)dv+A_{\text{diff}},
      \end{align}
      where $A_{\text{diff}}\geq0$. Therefore, we can write
      \begin{align}\nonumber
      L(f,g,\lambda)&=\frac{1}{\sigma_v}\left(\int\limits_{(I_1\cup I_2)^{C}}\Phi\left(\frac{v}{\sigma_v}\right)dv\right.\\
      &\left.\quad+C(u)\cdot\int\limits_{  I_{1} \cup  I_{2}}\Phi\left(\frac{v}{\sigma_v}\right)dv\right)\\\nonumber
      &=\frac{1}{\sigma_v}\left(\int\limits_{(I_1^s\cup I_2^s)^{C}}\Phi\left(\frac{v}{\sigma_v}\right)dv+A_{\text{diff}}\right.\\
      &\left.\quad+C(u)\!\cdot\! \left(\int\limits_{  I_{1}^s \cup  I_{2}^s}\!\Phi\!\left(\frac{v}{\sigma_v}\right)\!dv\!-\!A_{\text{diff}}\right)\right)\\\nonumber
      &=\frac{1}{\sigma_v}\left(\int\limits_{(I_1^s\cup I_2^s)^{C}}\Phi\left(\frac{v}{\sigma_v}\right)dv\right.\\
      &\left.\quad+C(u)\!\cdot\!\int\limits_{  I_{1}^s \cup  I_{2}^s}\!\Phi\left(\frac{v}{\sigma_v}\right)\!dv\!\right)\!+\!A_{\text{diff}}(1\!-\!C(u))\\
      &=L^s(f,g,\lambda)+A_{\text{diff}}(1-C(u)),
      \end{align}
      where $L^s(f,g,\lambda)$ represents the Lagrangian for symmetric non-overlapping intervals. Since $A_{\text{diff}}(1-C(u))\geq 0$, it follows that
      \begin{align}\nonumber
      L(f,g,\lambda)\geq L^s(f,g,\lambda),
      \end{align}
      proving the claim.

Focusing on symmetric intervals, we define the positive parameter $0\leq a\leq \sqrt{D}$ such that $I_{2}=[-2\sqrt{D}+a,a]$ and $I_{1}=[-a,2\sqrt{D}-a]$, and $I_1\cap I_2=[-a,a]$. Therefore, (\ref{Ch2_Eqn:52}) can be rewritten as
\begin{align}\nonumber
L(f,g,\lambda)&=2Q\left(\frac{2\sqrt{D}-a}{\sigma_v}\right)+2\left(Q\left(\frac{u}{\sigma_w}\right)+\lambda u^2\right)\\\label{Ch2_Eqn:53}
&\quad\cdot
\left(Q\left(\frac{a}{\sigma_v}\right)-Q\left(\frac{2\sqrt{D}-a}{\sigma_v}\right)\right).
\end{align}
Optimizing (\ref{Ch2_Eqn:53}) over $a$ completes the proof of Proposition \ref{Ch2_Result:4}.

\QEDA

\section{Proof of Proposition \ref{Ch2_Result:5} } \label{Ch2_Appendix:5}
\textit{Necessary optimality condition of an encoder mapping $f$ for a given set of reconstruction points $\{\hat{v}_{(1)},\ldots,\hat{v}_{(K)}\}$}: We start by expanding $\epsilon(D)$ with respect to the different intervals defined in (\ref{Ch2_Eqn:94}). We have
\begin{subequations}
\begin{align}
&\epsilon(D)=1-\mathrm{Pr}\left((V-\hat{V})^2 < D\right)\\
&=1-\!\!\!\!\!\!\sum\limits_{\substack{{\mathcal{V}} :\\ {\mathcal{V}} \subset\left\{\hat{v}_{(1)},...,\hat{v}_{(K)}\right\}}}\!\!\!\!\!\!\mathrm{Pr}\left((V-\hat{V})^2< D|V\in  I_{\mathcal{V}}\right)\mathrm{Pr}\left(V\in  I_{\mathcal{V}}\right)\\\nonumber
&=1-\mathrm{Pr}\left((V-\hat{V})^2 < D|V\in  I_\emptyset\right)\mathrm{Pr}\left(V\in  I_\emptyset\right)\\\nonumber
&\quad-\mathrm{Pr}\left((V-\hat{V})^2 < D|V\in  I_{\left\{\hat{v}_{(1)},\ldots,\hat{v}_{(K)}\right\}}\right)\\\nonumber
&\quad\quad\cdot\mathrm{Pr}\left(V\in I_{\left\{\hat{v}_{(1)},\ldots,\hat{v}_{(K)}\right\}}\right)\\
&\quad-\!\!\!\!\!\sum\limits_{\substack{{\mathcal{V}} : {\mathcal{V}} \subset\left\{\hat{v}_{(1)},...,\hat{v}_{(K)}\right\}\\ |{\mathcal{V}}|=1,\ldots,K-1}}\!\!\!\!\!\mathrm{Pr}\left((V-\hat{V})^2< D|V\in  I_{\mathcal{V}}\right)\mathrm{Pr}\left(V\in  I_{\mathcal{V}}\right)\\\nonumber
&=1-\mathrm{Pr}\left(V\in  I_{\left\{\hat{v}_{(1)},\ldots,\hat{v}_{(K)}\right\}}\right)\\
&\quad-\!\!\!\!\!\sum\limits_{\substack{{\mathcal{V}} : {\mathcal{V}} \subset\left\{\hat{v}_{(1)},...,\hat{v}_{(K)}\right\}\\ |{\mathcal{V}}|=1,\ldots,K-1}}\!\!\!\!\!\mathrm{Pr}\left((V-\hat{V})^2< D|V\in  I_{\mathcal{V}}\right)\mathrm{Pr}\left(V\in  I_{\mathcal{V}}\right)\\\nonumber
&=1-\mathrm{Pr}\left(V\in  I_{\left\{\hat{v}_{(1)},\ldots,\hat{v}_{(K)}\right\}}\right)\\
&\quad-\!\!\!\!\!\sum\limits_{\substack{{\mathcal{V}} : {\mathcal{V}} \subset\left\{\hat{v}_{(1)},...,\hat{v}_{(K)}\right\}\\ |{\mathcal{V}}|=1,\ldots,K-1}}\!\!\!\frac{1}{\sigma_v}\int\limits_{v\in I_{\mathcal{V}}}\Phi\left(\frac{v}{\sigma_v}\right)\mathrm{Pr}\left(W+f(v)\in\zeta_{\mathcal{V}}\right)dv\\\nonumber
&=1-\mathrm{Pr}\left(V\in I_{\left\{\hat{v}_{(1)},\ldots,\hat{v}_{(K)}\right\}}\right)\\\label{Ch2_Eqn:70}
&\quad-\frac{1}{\sigma_v}\!\sum\limits_{\substack{{\mathcal{V}} : {\mathcal{V}} \subset\left\{\hat{v}_{(1)},...,\hat{v}_{(K)}\right\}\\ |{\mathcal{V}}|=1,\ldots,K-1}}~\int\limits_{v\in I_{\mathcal{V}}}\Phi\left(\frac{v}{\sigma_v}\right)\\
&\quad\quad\cdot\!\sum\limits_{j:\hat{v}_{(j)}\in {\mathcal{V}}}\!\left(\!Q\!\left(\!\frac{z_{(j-1)}\!-\!f(v)}{\sigma_w}\!\right)\!-\!Q\!\left(\!\frac{z_{(j)}\!-\!f(v)}{\sigma_w}\!\right)\!\right)dv,
\end{align}
\end{subequations}
where $\zeta_{\mathcal{V}}\triangleq\bigcup_{j:\hat{v}_{(j)}\in {\mathcal{V}}}\left[z_{(j-1)},z_{(j)}\right)$. Note that the different sets in (\ref{Ch2_Eqn:70}) partition the real line, that is, they are disjoint and their union is the real line. Substituting (\ref{Ch2_Eqn:70}) in (\ref{Ch2_Eqn:9}), the Lagrangian $L(f,g,\lambda)$ can be written in the form of (\ref{Ch2_Eqn:27}) with $\tilde{F}(v,f)$ defined as
\begin{align}\nonumber
&\tilde{F}(v,f)\triangleq\\
&\left\{\begin{array}{ll}
                       1+\lambda f(v)^2 & v\in  I_{\emptyset}, \\
                       \lambda f(v)^2 & v\in  I_{\left\{\hat{v}_{(1)},\ldots,\hat{v}_{(K)}\right\}}, \\
                       1\!-\!\!\sum\limits_{j:\hat{v}_{(j)}\in {\mathcal{V}}}\!\left(\!Q\!\left(\!\frac{z_{(j-1)}\!-\!f(v)}{\sigma_w}\!\right)\!\right. & v\in I_{\mathcal{V}},~|{\mathcal{V}}|=1,\ldots,K-1.\\
                       \left.-Q\!\left(\!\frac{z_{(j)}\!-\!f(v)}{\sigma_w}\!\right)\!\right)\!+ \! \lambda f(v)^2&
                     \end{array}\right.
\end{align}
Writing the necessary condition in (\ref{Ch2_Eqn:28}) and setting to zero leads to (\ref{Ch2_Eqn:81}).

\textit{Optimal decoder function $g$ for a given encoder mapping $f$}: For a given encoder mapping $f$, the optimal decoder can be obtained as
\begin{subequations}
\begin{align}
\hat{v}_{(j)}&=\arg\min_{t}~\mathrm{Pr}((V-t)^2\geq D|Y=y_{(j)})\\
&\equiv\arg\max_{t}~\mathrm{Pr}\left((V-t)^2< D,Y=y_{(j)}\right)\\
&=\arg\max_{t}~\int\limits_{t-\sqrt{D}}^{t+\sqrt{D}}\int\limits_{z_{(j-1)}-f(v)}^{z_{(j)}-f(v)}\Phi\left(\frac{v}{\sigma_v}\right)\Phi\left(\frac{w}{\sigma_w}\right)dw dv\\\nonumber
&=\arg\max_{t}~\int\limits_{t-\sqrt{D}}^{t+\sqrt{D}}\Phi\left(\frac{v}{\sigma_v}\right)\left(Q\left(\frac{z_{(j-1)}-f(v)}{\sigma_w}\right)\right.\\\label{Ch2_Eqn:64}
&\left.\quad\quad\quad\quad-Q\left(\frac{z_{(j)}-f(v)}{\sigma_w}\right)\right)dv.
\end{align}
\end{subequations}

\QEDA

\section{Proof of Proposition \ref{Ch2_Result:6} } \label{Ch2_Appendix:6}
Define $y_i\triangleq \Gamma (f(v)+w_i)$, $w^N\triangleq[w_1,...,w_N]^T$, $y^N\triangleq[y_1,...,y_N]^T$. Expanding $L(f,g,\lambda)$ in (\ref{Ch2_Eqn:4}) for the MSE we have
\begin{subequations}
\begin{align}\nonumber
&L_s(f,g,\lambda)\\\nonumber
&=\frac{-1}{\sigma\sigma_v}\int\limits_{v}\int\limits_{w^N} v g\left(y^N\right)\Phi\left(\frac{v}{\sigma_v}\right)\prod\limits_{i=1}^{N}\Phi\left(\frac{w_i}{\sigma_{w_i}}\right)dw^Ndv\\
&\quad+\lambda \frac{1}{\sigma_v}\int \Phi\left(\frac{v}{\sigma_v}\right)f(v)^2 dv\\
&=\!\frac{-1}{\sigma\sigma_v}\!\int\! \Phi\!\left(\!\frac{v}{\sigma_v}\!\right)\!\left(\!v\!\int\limits_{w^N} \!g\left(y^N\right)\!\prod\limits_{i=1}^{N}\!\Phi\!\left(\!\frac{w_i}{\sigma_{w_i}}\!\right)\!dw^N\!-\!\lambda f(v)^2\!\right)\! dv\\\label{Ch2_Eqn:55}
&=\!\frac{-1}{\sigma\sigma_v}\!\int \! \Phi\!\left(\!\frac{v}{\sigma_v}\!\right)\! \! \left(\!v\sum\limits_{j=1}^{2^N}\hat{v}_{(j)}\mathrm{Pr}\!\left(\!\hat{V}=\hat{v}_{(j)}|V=v\!\right)\!-\! \lambda f(v)^2\!\right)\! dv,
\end{align}
\end{subequations}
where $\hat{v}_{(j)}=\mathbb{E}[V|Y^N=b_j^N]=\mathbb{E}[V|Y_1=b_{j}^N(1),\ldots,Y_N=b_{j}^N(N)]$ and $\sigma=\prod\limits_{i=1}^{N}\frac{1}{\sigma_{w_i}}$. We can compute
\begin{subequations}
\begin{align}
\mathrm{Pr}\left(\hat{V}=\hat{v}_{(j)}|V=v\right)&=\prod\limits_{i=1}^{N}\mathrm{Pr}\left( \Gamma (f(v)+W_i)=b_{j}^N(i)\right)\\\label{Ch2_Eqn:56}
&=\prod\limits_{i=1}^{N}Q\left(\frac{(-1)^{b_{j}^N(i)+1}f(v)}{\sigma_{w_i}}\right),
\end{align}
\end{subequations}
for $j=1,...,2^N$. Substituting (\ref{Ch2_Eqn:56}) in (\ref{Ch2_Eqn:55}), we have
\begin{align}\nonumber
L_s(f,g,\lambda)
&=\frac{-1}{\sigma_v}\int \Phi\left(\frac{v}{\sigma_v}\right)\left(v\sum\limits_{j=1}^{2^N}\hat{v}_{(j)}\right.\\\label{Ch2_Eqn:77}
&\left.\quad\cdot\prod\limits_{i=1}^{N}Q\left(\frac{(-1)^{b_{j}^N(i)+1}f(v)}{\sigma_{w_i}}\right) -\lambda f(v)^2 \vphantom{\sum\limits_{j=1}^{2^N}} \right)dv,
\end{align}
where for $\hat{v}_{(j)}$ we have
\begin{eqnarray}\label{Ch2_Eqn:57}
\hat{v}_{(j)}\!=\!\frac{\frac{1}{\sigma_v}\int v \Phi\left(\frac{v}{\sigma_v}\right)\prod\limits_{i=1}^{N}Q\left(\frac{(-1)^{b_{j}^N(i)+1}f(v)}{\sigma_{w_i}}\right)dv}{\frac{1}{\sigma_v}\int  \Phi\left(\frac{v}{\sigma_v}\right)\prod\limits_{i=1}^{N}Q\left(\frac{(-1)^{b_{j}^N(i)+1}f(v)}{\sigma_{w_i}}\right)dv}\!\triangleq\!\frac{r_{1_j}}{r_{2_j}},
\end{eqnarray}
for $j=1,\ldots,2^N$. Since (\ref{Ch2_Eqn:77}) is of the form (\ref{Ch2_Eqn:26}) with
\begin{align}
\tilde{F}&=-v\sum\limits_{j=1}^{2^N}\hat{v}_{(j)} \prod\limits_{i=1}^{N}Q\left(\frac{(-1)^{b_{j}^N(i)+1}f(v)}{\sigma_{w_i}}\right) +\lambda f(v)^2,\\
\tilde{G}_{1_j}&=v  \prod\limits_{i=1}^{N} Q\left(\frac{(-1)^{b_{j}^N(i)+1}f(v)}{\sigma_{w_i}}\right),~j=1,\ldots,2^N,\\
\tilde{G}_{2_j}&=\prod\limits_{i=1}^{N}Q\left(\frac{(-1)^{b_{j}^N(i)+1}f(v)}{\sigma_{w_i}}\right),~j=1,\ldots,2^N.
\end{align}
Writing down the optimality condition in (\ref{Ch2_Eqn:11}), we have
\begin{subequations}
\begin{align}\nonumber
\nabla L=&-v\sum\limits_{j=1}^{2^N}\frac{r_{1_j}}{r_{2_j}} \Theta \left(N,f(v),j\right)+2\lambda f(v)\\\nonumber
&-\sum\limits_{j=1}^{2^N}\frac{v\cdot\Theta \left(N,f(v),j\right)}{r_{2_j}}\cdot r_{1_j}\\
&+\sum\limits_{j=1}^{2^N}\frac{r_{1_j}\cdot\Theta \left(N,f(v),j\right)}{r_{2_j}^2}\cdot r_{1_j}\\
=&-\sum\limits_{j=1}^{2^N}\frac{r_{1_j}}{r_{2_j}} \Theta \left(N,f(v),j\right)\left(2v-\frac{r_{1_j}}{r_{2_j}}\right)+2\lambda f(v)\\
=&-\sum\limits_{j=1}^{2^N}\Theta \left(N,f(v),j\right)\hat{v}_{(j)} \left(2v-\hat{v}_{(j)}\right)+2\lambda f(v)=0,
\end{align}
\end{subequations}
where
\begin{align}\nonumber
\Theta \left(N,f(v),j\right)&=\sum\limits_{k=1}^{N}\left(\frac{(-1)^{b_{j}^N(k)}e^{-\frac{f(v)^2}{2\sigma_{w_k}^2}} }{\sqrt{2\pi}\sigma_{w_k}}\right.\\
&\left.\quad\quad\prod\limits_{l=1,l\neq k}^{N}Q\left(\frac{(-1)^{b_{j}^N(l)+1}f(v)}{\sigma_{w_l}}\right)\right).
\end{align}
Therefore, the optimal mapping must be in the form given by (\ref{Ch2_Eqn:37}).

\QEDA

\section{Proof of Proposition \ref{Ch2_Result:7} } \label{Ch2_Appendix:7}
\textit{Necessary optimality condition of an encoder mapping $f$ for a given set of reconstruction points $\left\{\hat{v}_{(1)},\ldots,\hat{v}_{(K)}\right\}$}: We expand $\epsilon(D)$
\begin{subequations}
\begin{align}
&\epsilon(D)=1-\mathrm{Pr}\left((V-\hat{V})^2< D\right)\\
&=1-\frac{1}{\sigma\sigma_v}\iint\limits_{v,w^N:|v-g(y^N)|^2< D}\Phi\left(\frac{v}{\sigma_v}\right)\prod\limits_{i=1}^{N}\Phi\left(\frac{w_i}{\sigma_{w_i}}\right) dw^Ndv\\\nonumber
&=1-\frac{1}{\sigma\sigma_v}\sum\limits_{{\mathcal{V}} : {\mathcal{V}} \subset\left\{\hat{v}_{(1)},...,\hat{v}_{\left(2^N\right)}\right\}}~~\int\limits_{v\in I_{\mathcal{V}}}\int\limits_{w^N: g(y^N)\in {\mathcal{V}} }\Phi\left(\frac{v}{\sigma_v}\right)\\
&\quad\quad\quad\quad\quad\quad\quad\quad\quad\quad\quad\quad\cdot\prod\limits_{i=1}^{N}\Phi\left(\frac{w_i}{\sigma_{w_i}}\right) dw^Ndv\\\nonumber
&=1-\frac{1}{\sigma_v}\sum\limits_{{\mathcal{V}} : {\mathcal{V}} \subset\left\{\hat{v}_{(1)},...,\hat{v}_{\left(2^N\right)}\right\}}~~\int\limits_{v\in I_{\mathcal{V}}}\Phi\left(\frac{v}{\sigma_v}\right)\\
&\quad\quad\cdot\sum\limits_{j: \hat{v}_{(j)}\in {\mathcal{V}} }\left(\prod\limits_{i=1}^{N}Q\left(\frac{(-1)^{b_{j}^N(i)+1}f(v)}{\sigma_{w_i}}\right)\right)dv\\\nonumber
&=1-\frac{1}{\sigma_v}\int\limits_{v\in I_{\left\{\hat{v}_{(1)},\ldots,\hat{v}_{\left(2^N\right)}\right\}}}\Phi\left(\frac{v}{\sigma_v}\right) dv\\\nonumber
&\quad-\frac{1}{\sigma_v}\sum\limits_{\substack{{\mathcal{V}} : {\mathcal{V}} \subset\left\{\hat{v}_{(1)},...,\hat{v}_{\left(2^N\right)}\right\}\\ |{\mathcal{V}}|=1,\ldots,2^N-1}}~~\int\limits_{v\in I_{\mathcal{V}}}\Phi\left(\frac{v}{\sigma_v}\right)\\\label{Ch2_Eqn:73}
&\quad\quad\quad\cdot\sum\limits_{j: \hat{v}_{(j)}\in {\mathcal{V}} }\left(\prod\limits_{i=1}^{N}Q\left(\frac{(-1)^{b_{j}^N(i)+1}f(v)}{\sigma_{w_i}}\right)\right)dv,
\end{align}
\end{subequations}
where $\sigma=\prod\limits_{i=1}^{N}\sigma_{w_i}$. Substituting (\ref{Ch2_Eqn:73}) in (\ref{Ch2_Eqn:4}), the Lagrangian $L(f,g,\lambda)$ can be written in the form of (\ref{Ch2_Eqn:27}) with $\tilde{F}(v,f)$ defined as

\begin{align}
&\tilde{F}(v,f)\triangleq\\
&\left\{\begin{array}{ll}
                       \!\!\!1+\lambda f(v)^2 & v\!\in\!  I_{\emptyset}, \\
                       \!\!\!\lambda f(v)^2 & v\!\in\!  I_{\left\{\hat{v}_{(1)},\ldots,\hat{v}_{\left(2^N\right)}\right\}}, \\
                       \!\!\!1+\lambda f(v)^2 &  v\!\in\! I_{\mathcal{V}}\!:\! |{\mathcal{V}}|\!=\!1,...,2^N\!-\!1.\\
                       \!\!\!-\!\!\!\sum\limits_{j: \hat{v}_{(j)}\in {\mathcal{V}} }\prod\limits_{i=1}^{N}\!Q\!\left(\!\frac{(-1)^{b_{j}^N(i)+1}f(v)}{\sigma_{w_i}}\!\right)&
                     \end{array}\right.
\end{align}
Writing the necessary condition in (\ref{Ch2_Eqn:28}) and setting to zero yields the result in (\ref{Ch2_Eqn:83}).

\textit{Optimal decoder function $g$ for a given encoder mapping $f$}: Assume that the encoder mapping is given. Following a similar approach to the derivation of the optimal decoder for a $K$-level ADC front end, the optimal decoder at the receiver is obtained as
\begin{subequations}
\begin{align}
\hat{v}_{(j)}\!&=\!\arg\max_{t}\!\int\limits_{t-\sqrt{D}}^{t+\sqrt{D}}\!
\!\Phi\!\left(\!\frac{v}{\sigma_v}\!\right)\!\mathrm{Pr}(Y_1\!=\!b_{j}^N(1),...,Y_N\!=\!b_{j}^N(N)) dv\\\label{Ch2_Eqn:63}
\!&=\!\arg\max_{t}\!\int\limits_{t-\sqrt{D}}^{t+\sqrt{D}}\!\Phi\!\left(\!\frac{v}{\sigma_v}\!\right)
\prod\limits_{i=1}^{N}Q\left(\frac{(-1)^{b_{j}^N(i)+1}f(v)}{\sigma_{w_i}}\right)dv.
\end{align}
\end{subequations}

\QEDA

\section{Proof of the decreasing characteristic of (\ref{Ch2_Eqn:41}) with respect to $|b|$ }\label{Ch2_Appendix:10}
It can be easily verified from (\ref{Ch2_Eqn:41}) that $\tilde{g}(b)$ is even. In the following, using the bounding techniques for the $Q$ function, we show that $\tilde{g}(b)$ is decreasing with respect to the absolute value of $b$. We assume $\sigma_v^2=1$ for brevity of the presentation, but the results are valid for any value of $\sigma_v^2$. We consider the cases $b\geq 1$ and $0\leq b \leq 1$ separately.

\begin{itemize}
\item $b\geq 1$: Taking the derivative of (\ref{Ch2_Eqn:41}) with respect to $b$, we have
\begin{align}\label{Eqn:42}
\tilde{g}^{'}(b)=\frac{e^{-b^2}g_1(b)}{2\pi\left(Q(b)\left(1-Q(b)\right)\right)^2}.
\end{align}
where $g_1(b)\triangleq\Phi(b)(1-2Q(b))-2bQ(b)(1-Q(b))$.

It can be easily verified that $\tilde{g}^{'}(0)=0$. It is enough to show that $g_1(b)<0$ for $b>1$. Using the lower bound $Q(b)\geq \frac{b}{1+b^2}\Phi(b)$, we have
\begin{align}
g_1(b)&\leq \Phi(b)(1-2Q(b))-\frac{2b^2}{1+b^2}\Phi(b)(1-Q(b))\\
& =\frac{\Phi(b)}{1+b^2}\left(1-b^2-2Q(b)\right),
\end{align}
which is obviously negative for $b>1$.

\item $0\leq b \leq 1$: Taking the second derivative of $g_1(b)$, we have
\begin{align}
\tilde{g}^{''}_1(b)=-\Phi(b)\left((b^2-3)(1-2Q(b))+2b\Phi(b)\right).
\end{align}
Using the first two terms of the Taylor series expansion of the $Q$-function as a lower bound, that is, $Q(b)\geq 1/2-\Phi(0)b$, we can write
\begin{align}
\tilde{g}^{''}_1(b)&\geq -\Phi(b)\left((b^2-3)2\Phi(0)b+2b\Phi(0)\right)\\
&\geq -2b \Phi(0) \Phi(b) (b^2-2)\geq 0.
\end{align}
Therefore, $g_1(b)$ is convex for $0\leq b\leq 1$. Noticing that $g_1(0)=0$, and $g_1(1)<0$, concludes that $g_1(b)<0$ for all $b>0$.
\end{itemize}

\QEDA

\bibliographystyle{ieeetran}
\bibliography{ref}

\begin{IEEEbiographynophoto}
{Morteza Varasteh}
[S'13]
received the B.S. degree in electrical and electronics engineering from Tabriz University, Tabriz, Iran, in 2009, and the M.S. degree in communication systems engineering from Sharif University of Technology, Tehran, Iran, in 2011, and the Ph.D. degree in communications from Imperial College London, London, UK, in 2016. He is currently a postdoctoral research associate in the Communications and Signal Processing group at Imperial College London. His research interests are in the general areas of information theory, wireless communications, and optimization theory.
\end{IEEEbiographynophoto}

\begin{IEEEbiographynophoto}
{Borzoo Rassouli}
[S'13]
received his M.Sc. degree in communication systems engineering from University of Tehran, Iran, in 2012, and Ph.D. degree from Imperial College London in 2016. He is currently a postdoctoral researcher in the Intelligent Systems and Networks Group at Imperial College. His research interests are in the general areas of information theory, wireless communications, and detection and estimation theory.
\end{IEEEbiographynophoto}

\begin{IEEEbiographynophoto}
{Osvaldo Simeone}
[SM'14] is a Professor of Information Engineering with the Department of Informatics at King's College London. He received an M.Sc. degree (with honors) and a Ph.D. degree in information engineering from Politecnico di Milano, Milan, Italy, in 2001 and 2005, respectively. He was previously a Professor affiliated with the Center for Wireless Information Processing (CWiP) at the New Jersey Institute of Technology (NJIT). His research interests include wireless communications, information theory, optimization and machine learning. Dr Simeone is a co-recipient of the 2017 JCN Best Paper Award, the 2015 IEEE Communication Society Best Tutorial Paper Award and of the Best Paper Awards of IEEE SPAWC 2007 and IEEE WRECOM 2007. He was awarded a Consolidator grant by the European Research Council (ERC) in 2016. His research has been supported by the U.S. NSF, the ERC, the Vienna Science and Technology Fund, as well by a number of industrial collaborations. Dr Simeone is a co-author of a monograph, an edited book published by Cambridge University Press and more than one hundred research journal papers. He is currently a Distinguished Lecturer of the Information Theory Society and a member of the Signal Processing for Communications and Networking Technical Committee. He is a Fellow of the IEEE.
\end{IEEEbiographynophoto}

\begin{IEEEbiographynophoto}
{Deniz G\"{u}nd\"{u}z}
[S'03-M'08-SM'13] received the B.S. degree in electrical and electronics engineering from METU, Turkey in 2002, and the M.S. and Ph.D. degrees in electrical engineering from NYU Polytechnic School of Engineering in 2004 and 2007, respectively. After his PhD, he served as a postdoctoral research associate at Princeton University, and as a consulting assistant professor at Stanford University. He was a research associate at CTTC in Barcelona, Spain until September 2012, when he joined the Electrical and Electronic Engineering Department of Imperial College London, UK, where he is currently a Reader in information theory and communications.

His research interests lie in the areas of communications and information theory with special emphasis on wireless networks, joint source-channel coding, and privacy in cyber-physical systems. Dr. Gündüz is an Editor of the IEEE TRANSACTIONS ON COMMUNICATIONS, and the IEEE TRANSACTIONS ON GREEN COMMUNICATIONS AND NETWORKING. He is the recipient of a Starting Grant of the European Research Council (ERC) in 2016, IEEE Communications Society Best Young Researcher Award for the Europe, Middle East, and Africa Region in 2014, Best Paper Award at the 2016 IEEE Wireless Communications and Networking Conference (WCNC), and the Best Student Paper Award at the 2007 IEEE International Symposium on Information Theory (ISIT). He served as the General Co-chair of the 2016 IEEE Information Theory Workshop, and a Co-chair of the PHY and Fundamentals Track of the 2017 IEEE Wireless Communications and Networking Conference.
\end{IEEEbiographynophoto}

\end{document}